\documentclass[final]{siamltex}

\usepackage[]{fontenc}
\usepackage[utf8]{inputenc}
\usepackage{verbatim}
\usepackage{amsmath}
\usepackage{amssymb}
\usepackage{dsfont}
\usepackage{amsfonts}
\usepackage{mathrsfs}
\usepackage{graphicx}
\usepackage{color}
\usepackage{ulem}
\usepackage{xspace,bbm,epic,eepic}
\usepackage{pgfplots}
\usepackage{setspace}
\usepackage{xspace}

\usepackage{stmaryrd}
\usepackage{graphicx}
\usepackage{lineno}
\usepackage{color}
\usepackage{url}
\usepackage{algorithm}
\usepackage{algpseudocode}

\usepackage{amsmath,amssymb,color} 
\usepackage{mathtools}

\usepackage{hyperref}

\newcommand{\Space}{\mathcal{X}}

\newcommand{\mref}{\pi}
\newcommand{\Score}{S^{\ast}}
\newcommand{\Srom}{S}
\newcommand{\Err}{E}

\newcommand{\mrm}[1]{\mathrm{#1}}

\newcommand{\tit}[1]{\textit{#1}}

\newcommand{\Ent}{\mathrm{Ent}}

\newcommand{\eg}{\textit{e.g.}, }
\newcommand{\ie}{\textit{i.e.}, }

\DeclareMathOperator{\Var}{\mathbb{V}ar}

\newcommand{\bigO}{\mathcal{O}}

\newcommand{\E}{\mathbb{E}}

\newcommand{\R}{\mathbb{R}}

\newcommand{\e}{\mathrm{e}}
\renewcommand{\d}{\mathrm{d}}

\newcommand{\eqdef}{ \dps \mathop{=}^{{\rm def}} }
\newcommand{\one}{ \, {\rm l} \hspace{-.7 mm} {\rm l}}
\newcommand{\dps}{\displaystyle}

\newcommand{\abs}[1]{\left | #1\right |}
\newcommand{\set}[1]{\left\{#1\right\}}
\newcommand{\p}[1]{ \left(#1\right) }
\renewcommand{\b}[1]{\left [ #1\right ]}

\newtheorem{The}{Theorem}[section]

\newtheorem{Pro}[The]{Proposition}

\numberwithin{equation}{section}

\begin{document}

\title{Adaptive Reduced Tempering For Bayesian inverse problems and rare event simulation}

\author{Fr\'ed\'eric C\'erou$^{*}$, Patrick H\'eas$^{*}$,   Mathias Rousset\thanks{Inria and Irmar,  University of Rennes, France. \\({\tt frederic.cerou@inria.fr,patrick.heas@inria.fr,mathias.rousset@inria.fr}) }
   }

\date{}
\maketitle

\begin{abstract}
This work proposes an adaptive sequential Monte Carlo sampling algorithm to solve Bayesian inverse problems in scenarios where likelihood evaluations are costly but can be approximated using a surrogate model built from previous evaluations of the true likelihood.  A rough estimate of the surrogate error is required. The method relies on an adaptive SMC framework that simultaneously adjusts both the likelihood approximations and a standard tempering scheme of the target posterior distribution.  This algorithm is well-suited for cases where the posterior is concentrated in a rare and unknown region of the prior.  It is also suitable for solving low-temperature and rare event simulation problems. The main contribution is to propose an entropy criterion that relates the accuracy of the current surrogate to a maximum inverse temperature for the likelihood approximation. The latter is instrumental to sample a so-called snapshot, on which is performed an exact likelihood evaluation, used to update the surrogate and its error quantification. Some consistency results are presented in an idealized framework for the proposed algorithm. Our numerical experiments use in particular a reduced basis approach to construct approximate parametric solutions to a partially observed solution of an elliptic partial differential equation. They demonstrate the convergence of the algorithm and show a significant cost reduction (close to a factor of $10$) for comparable accuracy.
\end{abstract}
\begin{keywords}
Bayesian sampling, reduced modeling, importance sampling, sequential Monte Carlo, adaptive tempering,  inverse problems, low-temperature simulation, rare events, relative entropy.  \vspace{-0.25cm}
\end{keywords}

\section{Introduction}

\subsection{Context and problem}\label{sec:contextProblem}
We consider the classical problem of designing an algorithmic Monte Carlo procedure that samples {from} a target probability distribution $\eta^\ast_{\beta_\infty}$ (for one specific $\beta_\infty \in \mathbb{R}$) defined explicitly up to a normalization {as} an element of the family
\begin{equation}\label{eq:target}
\eta^\ast_\beta \eqdef \frac{1}{Z^\ast_\beta} e^{\beta \Score} \d \mref, \qquad \beta \in \mathbb{R}.
\end{equation}
In the above, the quantity $Z^\ast_\beta \eqdef \mref(  e^{\beta \Score} )$ denotes the associated unknown normalization which typically must also be computed by the Monte Carlo procedure. $\mref$ denotes a given, easily simulable, reference probability distribution over a state space $\Space$, and $\Score:\Space  \to \mathbb{R}$ is a given but expensive-to-compute real-valued \textit{score} function.

Two motivating contexts are considered, where the score function takes the form:
\[ \Score(x) = \mathrm{score}(\Psi^\ast(x)), \]
where $\mathrm{score}$ is an easily computable function of physical interest depending on some additional parameters (like observed data), and $\Psi^\ast(x)$ is the outcome of a numerical computation of a complex physical system parametrized by the variable $x \in \Space$.

\begin{itemize}
\item In the \emph{Bayesian formulation of inverse problems}, $\beta_\infty = 1$ and $\eta^\ast_{\beta_\infty}$ is the posterior distribution of some uncertainty parameters $x \in \Space$  in a partially observed physical system $\Psi^\ast(x)$. $\pi$ models the prior distribution of the uncertainty parameters and the weight $\propto e^{\Score}$ is the likelihood of those parameters given some partial observations. As usual, one wants to Monte Carlo sample and then approximate expectations {with respect to}  this posterior.

\item In \emph{rare events problems}, one wants to estimate the small probability $p^\ast \eqdef \mref(\{\Score \geq 1\})$ corresponding to a large prescribed level $\ell \in \mathbb{R}$, with a score defined as $\Score(x) = 1 - \max(\ell - \Psi^\ast(x), 0)/\ell$. As $\lim_{\beta_\infty \to \infty} \pi(e^{\beta_\infty \Score}) = p^\ast$, the estimation of the rare event probability turns out to be the estimation of the normalization constant $Z^\ast_{\beta_\infty}$ as $\beta_\infty \to \infty$, \ie as the temperature $1/\beta_\infty$ tends to zero.
\end{itemize}

 In this work, we will be interested in scenarios in which the practitioner has to face {two} types of difficulty.
The first difficulty is classical  and arises when $\beta$ increases (or when the measurement noise becomes small in a Bayesian context), yielding a low-temperature  problem, similar to global optimization. It  confronts the practitioner with a computational challenge:  the density defining $\eta^\ast_\beta$ concentrates the majority of the probability mass in specific but possibly {unknown} areas of $\Space$, which are roughly described by the maxima of $\Score$. Direct sampling with a Monte Carlo simulation of $\pi$  is often time-consuming or even infeasible, as it can require a  very large sample size. For instance to obtain a reasonable estimation variance,  the sample size should be of order of  $1/p^\ast$ in the rare event simulation context, which is obviously prohibitive. A very popular general strategy to simulate a sample according to $\eta^\ast_\beta$  is to resort to a {\it Sequential Monte Carlo} (SMC) strategy (see~\cite{del2006sequential}) that starts with a Monte Carlo sample of particles distributed according to $\mref$ and then sequentially samples $\eta^\ast_\beta$ for increasing values of $\beta$ by resorting to a combination of Importance Sampling (IS), re-sampling (selection) of weighted samples and mutations of particles based on suitable Markov Chain Monte Carlo transitions. More specifically, we will be interested in the adaptive variant of tempering~\cite{Beskos2016}. 

The second difficulty appears when {the numerical evaluation of $\Psi^\ast$ is extremely expensive}; so expensive that the number of evaluations required in, say, a usual SMC approach becomes prohibitive. In order to circumvent this issue, we assume that the physical model $\Psi^\ast$ can be approximated by a \textit{reduced model}  denoted $\Psi$; the cost of evaluating $\Psi$  being small as compared to one evaluation of the true score with $\Psi^\ast$. The reduced model $\Psi$ we consider is assumed to come with two key features. To explain these, let us denote from now on the approximate score function by
$$
\Srom(x) = \mrm{score}\p{\Psi(x)} \in \R.
$$

The first feature is an \textit{a posteriori} point-wise error quantification associated with $\Srom$. This error quantification comes in the form of  function 
$
\Err(x) 
$
 satisfying an approximate point-wise estimate of the form
 \begin{equation}\label{eq:error_prior}
 	\abs{\Srom(x) - \Score(x)} \approx  \Err(x),\quad \forall x \in \Space.
 \end{equation}
 We underline that the point-wise estimate \eqref{eq:error_prior} need not to be a very precise estimate on the error, but only a rough estimate revealing the trend of the error.

The second key feature assumes that we are able to {update the reduced model}, through a procedure of the following form
\begin{equation*}
 \set{(X_1,\Psi^\ast(X_1)),\ldots,(X_k,\Psi^\ast(X_k)} \xmapsto{\textrm{reduced model}}
  \Psi= \Psi^{(X_1,\ldots,X_k)}, 
\end{equation*}
which takes as an input any sequence of states in $\Space$, called here \textit{a sample of snapshots}, together with the solution of the associated true model $\Psi^\ast$ evaluated for each of those snapshots. The output is the reduced model $\Psi$. 
A reduced model can be constructed by various, potentially very different, means. In this work, we will consider any reduced modeling procedure and any approximate error quantification of the form~\eqref{eq:error_prior}.  When $\Psi^\ast(x)$ is the solution of a Partial Differential Equation (PDE) parametrized by the uncertainty variable $x$, a  model reduction procedure particularly suited to this context is {\it reduced basis} \cite{quarteroni2015reduced}.

The associated error estimate in \eqref{eq:error_prior} is updated simultaneously, and for reasons that will be explained in Section~\ref{sec:prop_temp}, we additionally require that the error  $\Err_1,\ldots,\Err_k$ vanishes on snapshot points $X_{1},\ldots,X_k$ already evaluated at iteration $k$ :
\begin{equation}\label{eq:error_vanish}
E^{(X_1,\ldots, X_k)}(X_j) = 0, \quad \forall j\le k.
\end{equation}

In order to improve the readability, and throughout the paper, we may use the superscript $(k)$ instead of $(X_1,\ldots,X_k)$, or even drop it for any quantity that depends on  the sample of snapshots  $(X_1,\ldots,X_k)$. For instance the reduced score constructed out of $k$ snapshots may be denoted:
\[
S(x) \eqdef S^{(k)}(x) \eqdef S^{(X_1,\ldots,X_k)}(x).
\]

\subsection{Previous works}\label{sec:previous}
{ 
Advanced  methods for Bayesian sampling of \eqref{eq:target} are gradient-based  {\it Markov chain Monte Carlo} (MCMC)  methods such as {\it Metropolis adjusted Langevin algorithm} or  {\it Hamiltonian Monte Carlo} \cite{neal2011mcmc} and {\it preconditioning} (of gradients)  techniques \cite{MCMCstuart2013}.  Such  MCMC methods are known to scale   with dimensionality in certain  contexts (see, for example,~\cite{dwivedi2019log,neal2011mcmc,bandeira2023free,green2015bayesian})  and may solve large-scale Bayesian  problems (see, for example,~\cite{bui2014solving,heas2023chilled}).
Sequential Monte Carlo (SMC) methods or other methods based on interacting clones/particles are an interesting alternative for Bayesian sampling.  These methods  have also shown to be stable  in certain high-dimensional context~\cite{Beskos2014}, in particular {\it adaptive SMC} simulation \cite{Beskos2016}, and have been employed to solve large-scale  Bayesian inverse problems (see, for example,~\cite{beskos2015sequential,kantas2014sequential}). 
 However, in a general setting,  it is difficult to argue theoretically for the superiority  of one sampler over another. From an empirical point of view, studies tend to show that, while both methods achieve similar outcomes when the optimal solutions are not  in the tail of the specified prior distribution,  SMC's strength lies in the
dispersion of its interacting particles compared to a single point in  MCMC sampling~\cite{jeremiah2011bayesian}. 
Let us mention that other sampling methods such as {\it parallel tempering} have been proposed  to accelerate MCMC convergence \cite{latz2021generalized}.
 Nevertheless, sampling with any of these methods becomes difficult in the present context, where only a too small  budget of evaluations of the score function $\Score$ is affordable.\medskip
 
In recent years, numerous MCMC and SMC methods based on reduced score $\Srom$ approximating $\Score$ have been studied to reduce the simulation cost. Among MCMC approaches for inverse Bayesian problems,  as expected,  convergence results  have been obtained  for a sequence of reduced scores converging to $\Score$~\cite{conrad2018parallel}. Refinements of this approach involve integrating an iterative update of the $\Score-\Srom$ approximation error into a reduced MCMC sampling scheme using $\Srom$,  given a predetermined estimate of the error  (based on the evaluation of the error for a set of samples drawn from the prior)~\cite{calvetti2018iterative}. It should be noted, however, that a priori sampling to build the reduced model, characterize the approximation error or do both is probably not a relevant approach in our context. Indeed, in  Bayesian  inverse problems with informative measurements or in low-temperature problems, the target distribution usually concentrates its mass in a small and specific but unknown region of the prior support. More adapted to this context is the strategy to adapt the reduced model during the simulation, using samples whose distribution gradually matches the final target distribution.

Adaptive construction of the reduced score embedded in a tempering scheme was suggested in \cite{li2014adaptive}. As mentioned above, such an adaptive construction of the approximation is relevant to our context, but this work remains quite specific as it relies on parametric Gaussian proposals designed by a cross-entropy method~\cite{rubinstein2004cross}.  Moreover, the adaptive construction of reduced score does not use possible error estimates.

Regarding SMC approaches, {\it multilevel  SMC} has been proposed to optimize the product cost $\times$  variance of the estimates; in the latter `multilevel' refers to a sequence of approximations  of the score function at different discretization levels~\cite{beskos2017multilevel}.  However, multilevel SMC requires precise characterization of the model-dependent variance bounds of the estimates, which is a tedious task. In addition, although it relies on a given hierarchy of reduced models or approximations, these are not updated online using snapshots.

 An adaptive SMC approach using a predetermined hierarchy of reduced scores has also been proposed in  \cite{latz2018multilevel}. This work tackles the tricky problem of selecting the right reduced model at the current temperature during simulation. Indeed, choosing a proper criterion to decide along the simulation when to refine the reduced score is quite an open question in general \cite{Gelman98}, although various heuristics have been proposed in the context or rare event simulation relying on some error quantification~\cite{bect2017bayesian,heas2020selecting,cerou2024adaptive}.
In their work, the authors of \cite{latz2018multilevel} propose a heuristic  consisting  of minimizing the variation of the particle weights over the two possible scenarios: refining or maintaining the reduced score. In this work, we go beyond these ideas and generally answer the question of when  to update the reduced score and with which snapshots. 

The authors of~\cite{latz2018multilevel} have also proposed an analogous method~\cite{wagner2020multilevel} in the context of rare event simulation, in which a predetermined hierarchy of reduced model is given. The authors of the present work have also proposed a method similar to the one studied in the present paper in  a rare event context: the {\it adaptive reduced multilevel\footnote{Note that the term `multilevel'  in ARMS refers to the level sets of the score function~\cite{cerou:hal-01417241}, while in `multilevel SMC'  this term refers to the different precision levels of the simulation~\cite{giles2015multilevel,beskos2017multilevel}} splitting} (ARMS) algorithm~\cite{cerou2024adaptive}. ARMS is based on a sequential simulation scheme very similar to SMC, which adapts a reduced score approximation while getting closer to the target rare event distribution. This method is nevertheless specific to the rare event simulation context in the sense that it is dedicated to sample targets being  the reference distribution conditionally to be in some rare event set. 
The method proposed in this paper is called adaptive reduced tempering (ART) and can be seen as a generalization of the ARMS algorithm to the case of general target distributions of the form of \eqref{eq:target}, which are relevant for Bayesian problems. However, as explained \textit{e.g.} in Section~\ref{sec:contextProblem}, the ART method is also suited to the context of rare event simulation, by considering the latter as a zero temperature limit of a Gibbs family.

\subsection{Contributions}

Compared to the pieces of work mentioned in Section~\ref{sec:previous}, the main novelty of the proposed ART algorithm is to propose an adaptive rule that triggers, within a SMC sampling method with tempered surrogate targets, \emph{when} (using an entropy criterion) and  \emph{where} (using an importance distribution and an  error quantification) to sample a new snapshot of the true model (serving to  update the reduced model). 

To do so, we consider the following exponential family of importance distributions
\begin{align}\label{eq:propa}
	\mu_\beta \eqdef \frac{1}{Z_\beta} \e^{\beta \Srom} \d \mref, \qquad \beta \in \R 
\end{align}
where the normalization 
$Z_\beta \eqdef \mref( \e^{\beta \Srom}  )$
represents an approximation of $Z^\ast_\beta$. An important idea is to interpret the latter  as {\it proposal distributions} (also called {\it importance distributions}), in the sense of importance sampling, of the target family \eqref{eq:target} . 
 In the context of  rare event simulation,   using  proposal distributions in this  exponential family rather than \tit{e.g.} in the form of indicator of events $\propto \one_{ \Srom \ge \ell } \d \mref$ with $\ell \in \mathbb{R}$ enabling  to get rid of the possible nesting conditions between successive proposals in the algorithm (see~\cite{cerou2024adaptive}). \medskip

The idea behind this work is to design an algorithm iterating on $k$ where $k$ is the number of newly sampled snapshots, and giving at iteration $k$ a well-chosen inverse temperature $\beta^{(k)}$ able to perform a reasonable importance sampling of the target distribution $\eta^\ast_{\beta^{(k)}}$ defined by \eqref{eq:target}. This adaptive importance sampling is performed with the sequence of proposals 
\[
\mu_{\beta^{(k)}} = \frac{1}{Z^{(k)}_{\beta^{(k)}}} \e^{\beta^{(k)} \Srom^{(k)}} \d \mref, \qquad k \geq 1
\] 
where the reduced score $\Srom^{(k)}$ is much cheaper to compute, and only the evaluation of the importance weight are expensive. Very importantly, as the iteration progresses, the proposal will adapt by adjusting the inverse temperature $\beta^{(k)}$ and refining the approximation of the reduced score $ \Srom^{(k)}$ using the history of the previous computations of the true model. In this framework, three issues emerge.

 A first important one is that one has  to address the definition of a criterion for setting the lowest inverse temperature  $\beta^{(k)}$ suitable for the accuracy of the current reduced score approximation $\Srom^{(k)}$. We recall the definition of the relative entropy {or Kullback-Leibler divergence between probabilities defined by $$\Ent(\eta \mid \mu) \eqdef  \int \ln \frac{\d \eta}{\d \mu} \d \eta $$ when $\eta$ is dominated by $\mu$ and extended to $+\infty$ otherwise.} We propose to constrain  the relative entropy between a \tit{surrogate target} denoted $\check{\eta}_\beta$ and the proposal as follows (for a given threshold $c_1$):
 \begin{equation}\label{eq:entropy_simple}
\beta^{(k)} \text{ verifies } 
\Ent(\check{\eta}^{(k)}_{\beta^{(k)}} \mid \mu^{(k)}_{\beta^{(k)}})  \leq c_1 .
 \end{equation}
Relative entropy can be used to estimate the log of the cost of \emph{importance sampling} of the surrogate $\check{\eta}^{(k)}_{\beta^{(k)}}$ by the proposal $\mu^{(k)}_{\beta^{(k)}}$, as will be discussed in more detail in Section~\ref{sec:prop_temp}. This criteria ensures that the proposal can yield samples that are reasonably typical under the true target $\eta^\ast_{\beta^{(k)}}$.
 The surrogate target is constructed using the approximate error bound~\eqref{eq:error_prior} of the reduced score approximation as follows:
 \begin{equation}\label{eq:pessTarget}
\check{\eta}_\beta  \eqdef \frac{1}{\check Z_\beta} e^{\beta(\Srom-\Err) } \d \mref \quad \textrm{with} \quad \check Z_\beta \eqdef \mref(e^{\beta(\Srom-\Err) } ).
\end{equation}
The choice is motivated by the fact that $\check{\eta}_\beta$ is roughly speaking a version of $\mu_\beta$ where typical states with large expected errors (hence `far away' from the already sampled snapshots) are penalized.

 A second issue that we will address in this work is the tuning of the expensive snapshot sampling and its associated reduced modeling update. At iteration $k+1$ the snapshot $X_{k+1}$ will be sampled from the proposal $\mu^{(k)}_{\beta^{(k)}}$ which ensures thanks to~\eqref{eq:entropy_simple} a sampling in areas consistent with the true target distribution  $\eta^{\ast}_{\beta^{(k)}}$ at inverse temperature $\beta^{(k)}$. In the first phase of the algorithm, we sample snapshots with an additional strong penalty weight favoring candidate states with a larger error (hence requiring priority correction). In the final phase of the algorithm, we remove that penalty to obtain snapshot-based optimal importance sampling estimation.
 \begin{figure}[h!]
 	\begin{center}
 		\begin{tabular}{ccc}
 			\includegraphics[width=0.5\textwidth]{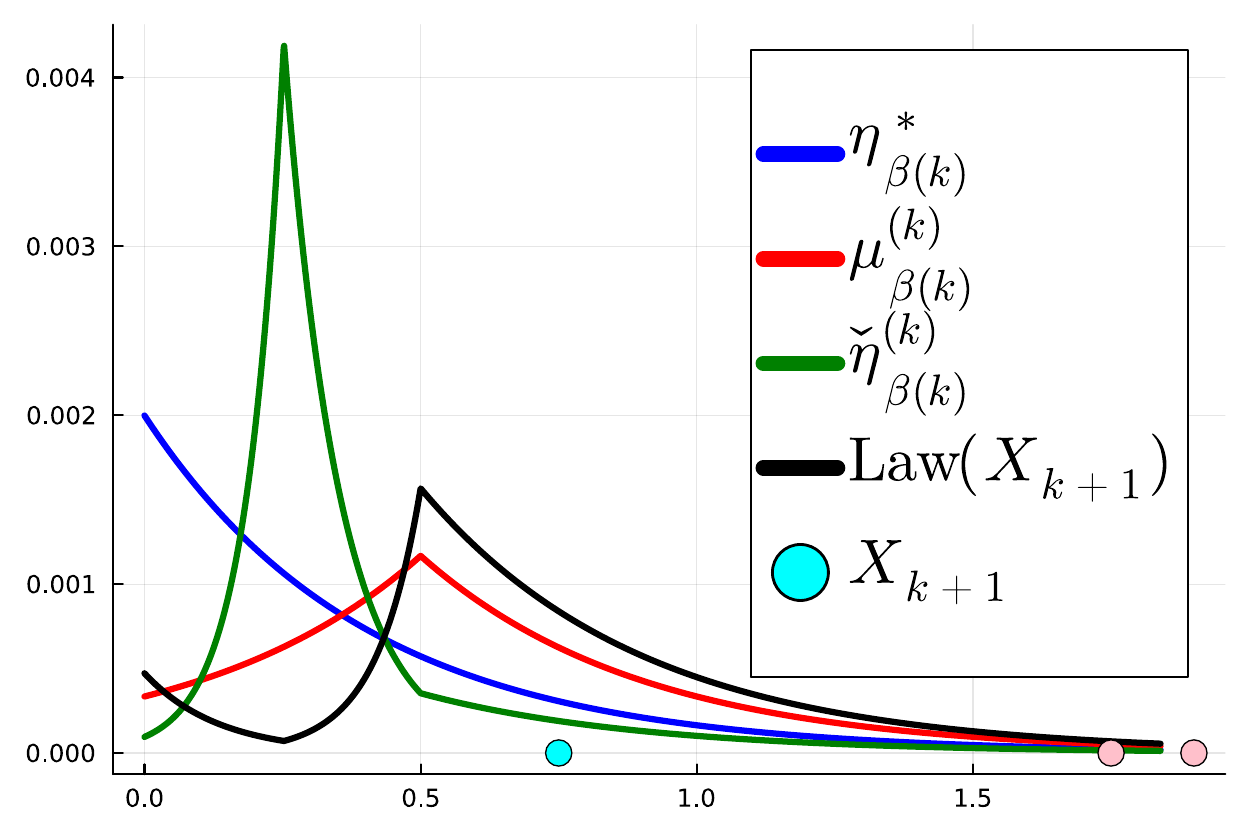}
 			\includegraphics[width=0.5\textwidth]{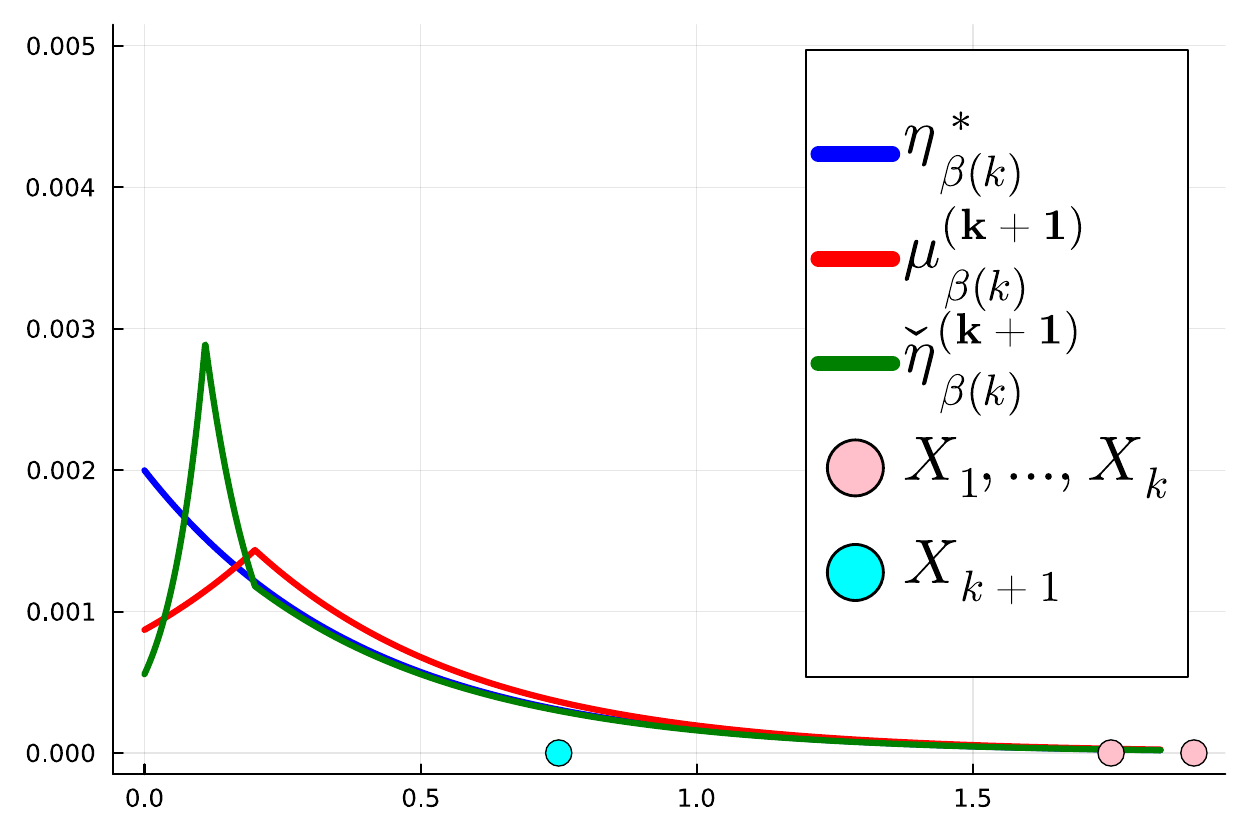} \\
 			\includegraphics[width=0.5\textwidth]{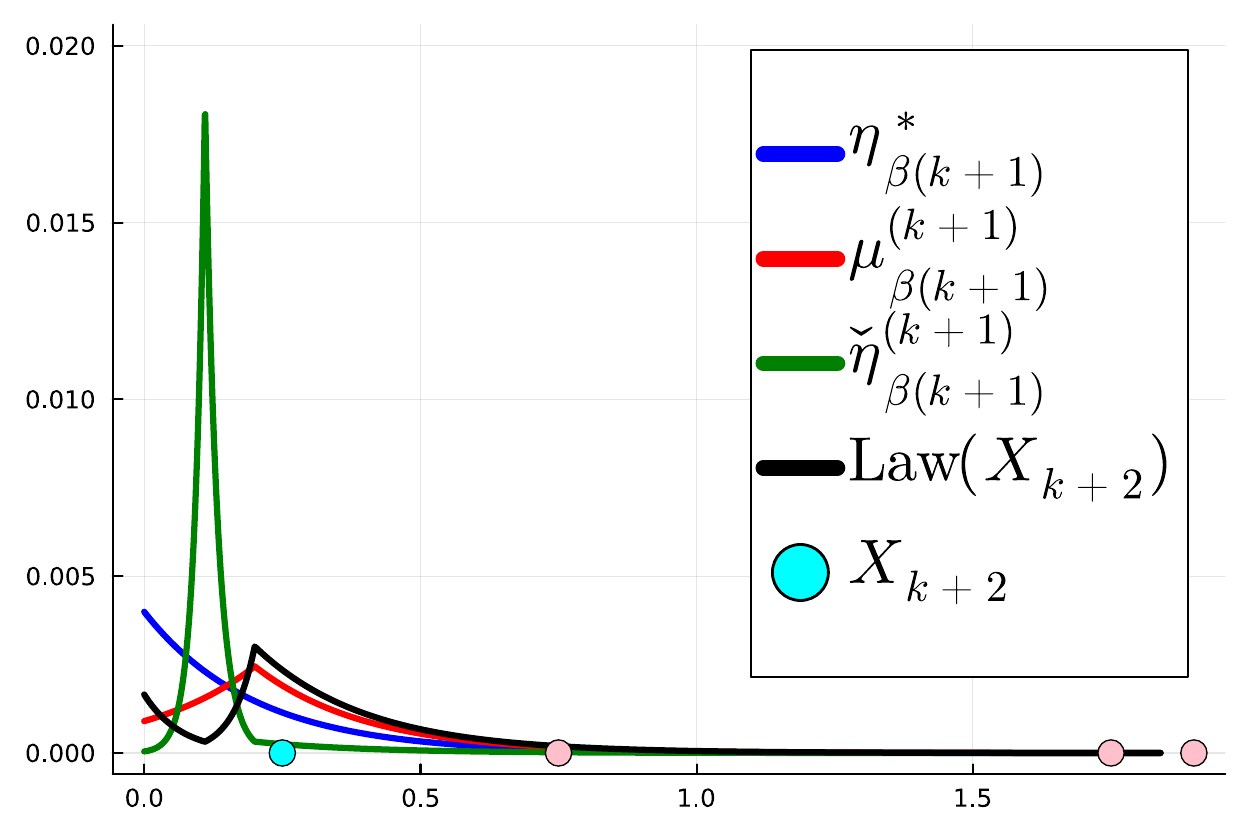}
 		\end{tabular}
 		\caption{{\footnotesize {\bf Illustration of one iteration of the main ART algorithm.} In the first picture, a snapshot $X_{k+1}$ (cyan) has been sampled with an error-weighted version (black) of the proposal $\mu^{(k)}_{\beta^{(k)}}$ (red). In the second picture, this proposal (red) is updated with the new snapshot using the reduced model. In the third picture, temperature is decreased until the cross-entropy between the updated proposal $\mu^{(k+1)}_{\beta}$ (red) and the surrogate target $\check \eta^{(k+1)}_{\beta}$ (green) reaches the small threshold $c_1$. Iteration follows. Note that the surrogate target is concentrated on the typical states of the proposal that have relatively small error. The entropic criteria ensures matching between target and proposal. }}\label{fig:intro}
 	\end{center}\vspace{-0.cm}
 \end{figure}

The main two successive steps: i) temperature tuning, and ii) new snapshot sampling, do form the core steps of our  methodology and are summarized in Fig.~\ref{fig:intro}. A corresponding moderately detailed idealized pseudo-code can be found in Section~\ref{subsec:pseudo} after more detailed descriptions of the various steps.

A more optional issue that has  been addressed in this work  is  the simulation of a new proposal when at the end of iteration $k$ the reduced score is updated $S^{(k)} \to S^{(k+1)}$. We will propose to {\it bridge}  proposals by searching among past simulations for an iteration index $k_{bridge}$ with associated approximate score $\Srom^{(k_{bridge})}$, and by finding an inverse temperature  $\beta_{bridge}$ such that the relative entropy between the new  proposal  $\mu_{\beta_{\mrm{bridge}}}^{(k+1)}$ and the past proposal ${  \mu_{\beta^{(k_{\mrm{bridge}})}}^{(k_{\mrm{bridge}})} }$ is kept below a certain standard threshold:
\begin{equation}\label{eq:bridge_simple}
	\beta_{\mrm{bridge}},k_{\mrm{bridge}} \text{ verifies } 
	\Ent(\mu_{\beta_{\mrm{bridge}}}^{(k+1)} \mid  \mu_{\beta^{(k_{\mrm{bridge}})}}^{(k_{\mrm{bridge}})}  )  \leq c_2 .
\end{equation}


The paper is organized as follows. An idealized version of the proposed algorithm, in which we assume the exact evaluation of expectations and the exact sampling of distributions,  is presented in Section~\ref{sec:ideal}  together with the  underlying methodological concepts. The proposed importance sampling  estimator is  shown to be consistent in such an idealized setting. We  then provide in Section~\ref{sec:practical} a practical implementation of this idealized algorithm using  SMC simulations. In Section~\ref{sec:num}, we exhibit by numerical experiments its practical convergence in the case of rare event simulation and inverse Bayesian problems.  Most importantly, by evaluating the computational complexity in relation to the estimation error, we show that the use of reduced modeling (as compared to similar methods with only true model evaluation) significantly reduces the simulation cost  while achieving a comparable accuracy.   The final section deals with conclusions.

\section{Idealized Adaptive Reduced Tempering}\label{sec:ideal}


We describe in this section the ingredients of an idealized adaptive reduced tempering (ART) algorithm that will lead to the consistent  estimation of the expectation $\eta^\ast_{\beta_\infty}(\varphi)$ for any generic test function~$\varphi$. 

\subsection{Iterative importance sampling}\label{sec:IS}
We consider a vanishing initial inverse temperature $\beta^{(0)}=0$, so that the initial  proposal distribution matches the reference distribution $\mu^{(0)}_{\beta^{(0)}} =\pi $. Using this initialization, we will design in Section~\ref{sec:prop_temp} an adaptive procedure which determines at each iteration $k \geq 1$ ($k$ being the number of snapshot evaluations  using the true score $\Score$)  some inverse temperature $\beta^{(k)}\in [0,\beta_\infty]$  ``suited'' to the current reduced score  $\Srom^{(k)}$, in the sense of the log cost of {importance sampling} defined in  \eqref{eq:entropy_simple}. To this temperature is associated  the target ${\eta}^\ast_{\beta^{(k)}}$ as defined by \eqref{eq:target} and the proposal distribution $\mu^{(k)}_{\beta^{(k)}} $ as defined by \eqref{eq:propa}. 
Given this adaptive procedure determining the sequence of inverse temperatures $\{\beta^{(k)}\}_{k}$, consider the idealized case where we can compute the normalizations $\{Z^{(k)}_{\beta^{(k)}}\}_k$ and where we can draw exactly according to the proposals $X_{k+1} \sim\mu^{(k)}_{\beta^{(k)}} $ in order to compute the sample of snapshots $ \set{(X_1,\Score(X_1)),(X_2,\Score(X_2)),\ldots } $. Then, an  {\it iterative importance sampling estimator} of the unnormalized target ${\gamma}^\ast_{\beta_\infty} \eqdef Z_{\beta_\infty}{\eta}^\ast_{\beta_\infty}$ 
 is given by  
\begin{align}\label{eq:unestimator}
  \hat \gamma^H_{\beta_\infty}(\varphi)=\frac{1}{H}\sum_{k=K_{j_0}+1}^{K_{j_0}+H} Z^{(k)}_{\beta^{(k)}}e^{\beta_\infty\Score(X_{k+1}) - \beta^{(k)}\Srom^{(k)}(X_{k+1})} \varphi(X_{k+1}),
    \end{align}
with  $H$ denoting the size of the sample used for estimation and where $K_{j_0}$  denotes a  random iteration index  whose definition will be given in due time. Appendix~\ref{app:martingal} shows the consistency of \eqref{eq:unestimator}. The  self-normalized estimate of the target ${\eta}^\ast_{\beta_\infty}$  is
\begin{align}\label{eq:estimator}
  \hat \eta^H_{\beta_\infty}(\varphi)=\frac{\hat \gamma^H_{\beta_\infty}(\varphi)}{\hat \gamma^H_{\beta_\infty}(\one)}.
  \end{align}

Therefore, to enable the evaluation of these estimates,   we need at iteration $k$  of the algorithm  to {\it i)} adapt the inverse temperature $\beta^{(k)}$, so that it is ``suited'' to the current reduced score function $\Srom^{(k)}$ (see Section~\ref{sec:prop_temp}), {\it ii)} compute efficiently the normalization $Z_{\beta^{(k)}}$ (see Section~\ref{sec:adaptempering}).

\subsection{Critical inverse temperature}\label{sec:prop_temp}

The idea is to tune at iteration $k$ the inverse temperature  $\beta^{(k)}$ with a criterion evaluating that surpassing this inverse temperature should require, in some sense, a better surrogate model, and thus new snapshots. The adjusted proposal $\mu^{(k)}_{\beta^{(k)}}$ will then be used to sample a new snapshot, and obtain an updated score function $\Srom^{(k+1)}$ for the next iteration $k+1$. In short, we are going to develop a heuristic criterion  to detect at which tempertaure a new snapshot is needed.  

This idea is developed further below. We want to perform some sequential algorithm to sample the distributions $\mu_\beta^{(k)}$ along an increasing sequence of inverse temperatures $\beta$ as high as possible, and sometimes possibly use importance sampling to sample according to the target $\eta^\ast_\beta$ or a variant of it. However, we do not want to waste resources for temperature that are ``unreasonably'' low because importance sampling becomes unfeasible due to the mismatch between proposal and target. As an idealized criterion giving achievable temperatures, we propose the {\it  cost of importance sampling} of the true target. 
As is thoroughly argued in~\cite{ChaDia18} (see also~\cite{cerou2022entropy,cerou2024adaptive}), the variance is in general a pessimistic quantification of importance sampling. For a target $\eta$ with proposal $\mu$, it has been proven in~\cite{ChaDia18} that the cost  of importance sampling -- in terms of sample size -- is roughly given (up to a reasonable tail condition on the log-density\footnote{{We stress that this tail condition is satisfied by the pair $(\check \eta_\beta,\mu_\beta)$ we use in practice because the log likelihood is bounded by the finite uniform norm of the error $E$.}} $\ln \frac{\d \eta}{\d \mu}$), by the exponential of the relative entropy  between $\eta$ and $\mu$, \ie
$
\e^{\Ent(\eta \mid \mu)},
$
where we recall that $\Ent(\eta \mid \mu)=  \eta(\ln\frac{\d\eta}{\d\mu})$.
We will thus  call $\Ent(\eta \mid \mu)$ the {log cost of importance sampling} of the target $\eta$ by the proposal $\mu$. 
 The first non-achievable temperature, hence a kind of critical temperature,  is thus given when $\Ent(\eta^\ast_{\beta} \mid \mu_{\beta}^{(k)})$  matches a certain given numerical threshold $c_1$. By doing so, we ensure that the snapshot sampled with respect to $\mu_\beta^{(k)}$ will not diverge too much from the target distribution $\eta^\ast_{\beta}$ at an appropriate inverse temperature $\beta$. 

In a practical context, the log cost $\Ent(\eta^\ast_{\beta} \mid \mu_{\beta}^{(k)})$ has to be estimated using some prior information on $\eta^\ast_{\beta}$, for instance using some form of error quantification. We propose to consider a surrogate  of the target $\eta^\ast_\beta$, denoted by  $\check{\eta}_{\beta}^{(k)}$, constructed  according to \eqref{eq:pessTarget} with the reduced score $\Srom^{(k)}$ and its error quantification $\Err^{(k)}$ of the form of  \eqref{eq:error_prior}. We stress that this choice of surrogate is merely conventional and other cases could be studied in future work. Our choice is simply a limit case\footnote{Taking $\Srom + E$ instead of $\Srom - E$ in the definition of the surrogate $\check{\eta}_{\beta}$ is a possibility but it seems to lead to less robust calculations because it favors states with high errors. We did not study in detail different options for surrogate targets in the present work.} of possible true score imposed by the approximate error bound~\eqref{eq:error_prior}. The critical temperature  will be estimated using the surrogate target by searching the  closest temperature such that the log cost $\Ent\p{\check{\eta}_{\beta}^{(k)} \mid \mu_{\beta}^{(k)} } $  matches the threshold $c_1$.  
 We emphasize that the irrelevant situation where relative entropy is uniformly zero is excluded by the constraint \eqref{eq:error_vanish}, which requires that the error  $\Err^{(k)}$ vanishes on snapshot points $X^{(k)}$ already evaluated at iteration $k$. This prohibits constant error quantification functions unless the error estimate $E$ is identically zero. We would like to point out that this condition is naturally satisfied by many reduced models, and in particular by reduced bases. However, this requirement is not a hard limit on the method, as the choice of the surrogate $\check{\eta}_{\beta}$ can be adapted in other cases.

Given an initial inverse temperature $\beta_0$, this  yields  the following rigorous definition of  the inverse of the  critical temperature:
\begin{equation}\label{eq:level}
 \beta^{(k)}(c_1,\beta_0) \eqdef \min \{ \beta  : \; \Ent\p{\check{\eta}_{\beta}^{(k)} \mid \mu_{\beta}^{(k)} }  >  c_1\,,\, \beta_0 \leq \beta  \leq \beta_\infty \},
\end{equation}
with the initial inverse temperature at $k=0$ set to zero, \ie $\beta^{(0)}(c_1,\beta_0)\eqdef 0.$ {Note that the relative entropy in~\eqref{eq:level} is not necessarily monotonic in $\beta$, by definition we simply pick the smallest possible $\beta \geq \beta_0$, if it exists}.
For convenience, the dependency on $c_1$ and $\beta_0$ of $\beta^{(k)}$ will be removed if the context makes this clear.
$\beta^{(k)}( c_1,\beta_0)$ is an inverse temperature such that, in the case where the true target is given by the surrogate $\check{\eta}_{l}^{(k)}$,  a  computational cost of ${\rm e}^c_1$  is approximately required in order to perform importance sampling.  
We will define $\beta_0$ later in  Section~\ref{sec:bridging}.

\subsection{Adaptive tempering down to the critical temperature}\label{sec:adaptempering}

In the previous section,  we have defined  $\beta^{(k)}$ as the critical level at the $k$-th iteration of the algorithm. However, the computation of the normalization $Z^{(k)}_{\beta^{(k)}}$ related to the distribution $\mu^{(k)}_{\beta^{(k)}} $ is often unaffordable by direct Monte Carlo simulation for a large $\beta^{(k)}$, as it corresponds to a difficult low-temperature problem.  A frequently invoked procedure  for these problems involves forming a sequence of tempered probability distributions $\mu^{(k)}_{\beta} $ among the family \eqref{eq:propa}, defined   with the reduced score $\Srom^{(k)}$ and parameters $\beta_0 \le \beta_1 \le \cdots \le \beta_n =\beta^{(k)} $ \cite{Beskos2016}. Assume  given an initial normalization constant $Z^{(k)}_{\beta_0}$ related to an initial distribution $\mu^{(k)}_{\beta_0}$.  It is straightforward to show that, using this sequence,  the sought normalization  can be decomposed into the product 
\begin{equation}\label{normalizationRec}
Z^{(k)}_{\beta^{(k)}}=Z^{(k)}_{\beta_0}\prod_{i=1}^{n-1} \mu^{(k)}_{\beta_i} (G_i), \quad G_i(x)=e^{(\beta_{i+1}-\beta_{i})\Srom{(k)}(x)},
\end{equation}
where, as we will see in Section~\ref{sec:practical}, the factor $\mu^{(k)}_{\beta_i}(G_i)$ can be recursively approximated using a SMC simulation. 
An adaptive version of this procedure, known as {\it adaptive tempering}, corresponds to a flow of distributions with a sequence of temperature parameters chosen adaptively. In this work,  the inverse temperature $\beta_{i+1}$ will be adapted  using the rule
\begin{equation}\label{eq:ATcrietrion}
 \beta_{i+1} := \min \{  \beta : \; \Ent\p{ \mu_{\beta}^{(k)} \mid \mu_{\beta_{i}}^{(k)} } >  c_2 \,,\,  \beta_i \leq \beta \leq \beta^{(k)}\},
 \end{equation}
where $c_2>0$ is some given prescribed level. This one dimensional problem is easily solved since $\beta \mapsto \Ent\p{ \mu_{\beta}^{(k)} \mid \mu_{\beta_{i}}^{(k)} }$ is increasing.\footnote{Some simple  calculation 
shows that 
$ \partial_{\beta} \Ent\p{ \mu_{\beta}^{(k)} \mid \mu_{\beta_{i}}^{(k)} }=(\beta-{\beta_{i}})\textrm{var}_{\ \mu_{\beta}^{(k)}}(  {\Srom^{(k)}})$.}
The entropic criterion used in \eqref{eq:ATcrietrion} may be viewed as an extension to smooth distributions of the criterion used in adaptive multilevel splitting~\cite{cerou2024adaptive}. However, other criteria are possible, see for example~\cite{Beskos2016}.

\subsection{Snapshot sampling}\label{sec:learn}


As defined  in~Section~\ref{sec:IS}, only the snapshots sampled at and after iteration $K_{j_0}+1$ are  used to build the importance sampling estimator $  \hat \gamma^H_{\beta_\infty}$ and   $\hat \eta^H_{\beta_\infty}$. The idea here is that the snapshots should be first used  to refine the reduced score  $\Srom^{(k)}$ around the regime of interest. The snapshots can be used for importance sampling once the reduced score is sufficiently accurate in the sense of a limited variance for importance sampling with  distribution $\mu^{(k)}_{\beta^{(k)}} $ of the target $\eta_{\beta^{(k)}}^\ast$. We choose to define the  {\it trigger index}    $K_{j_0}$ (random index independent of the future snapshots) as the first iteration index $k$ for which among the set of already computed critical inverse temperatures $\{\beta^{(k')},\, k'\le k\}$, at least $j_0$ of them are equal to the targeted inverse temperature  $\beta_\infty$, for some prescribed $j_0\ge 1$. This snapshot sampling approach corresponds to some extent to a particular setting of the so-called \textit{learning function} in the active learning literature.~\cite{moustapha2022active}. 

In summary,  in the first iterations new snapshots are sampled using a large penalty favoring high error bounds:  $$X_{k+1} \sim \frac{1}{Z^{(k)}_\tau} \e^{\tau E^{(k)}(x)} \mu_{\beta^{(k)}}^{(k)}(\d x)\quad \textrm{if}\quad k< K_{j_0}.$$ 
In practice in the SMC implementation, for not too large sample size $N$, we may pick $\tau$ large enough to be  equivalent to $+\infty$. Then, once the reduced score is related to a temperature close to the target one, new snapshots are sampled  according to the  importance distribution:  $$X_{k+1} \sim\mu^{(k)}_{\beta^{(k)}}\quad \textrm{if}\quad k\ge K_{j_0}.$$

\subsection{Bridging reduced scores}\label{sec:bridging}

Given an inverse temperature $\beta^{(k)}$ and an associated importance distribution $\mu_{\beta^{(k)}}^{(k)}$ related to the reduced score $S^{(k)}$, the problem is to determine a new initial inverse temperature denoted by  $\beta_{\mrm{bridge}}(k+1)$,  ``suited''   to the updated reduced score $S^{(k+1)}$ related  to the updated importance distribution $\mu_{\beta_{\mrm{bridge}}(k+1)}^{(k+1)}$. 
In order to set this temperature in consistency with the critical temperature constraint \eqref{eq:level} and the adaptive tempering constraint \eqref{eq:ATcrietrion}, it is natural to consider the problem  
\begin{equation}\label{eq:bridgeProblem0}
\text{Find: } \beta \in \mathcal{C}(k),
\end{equation}
 where 
\begin{equation}\label{eq:bridgeconst}
\mathcal{C}(k)\eqdef \{ \beta  : \; \Ent\p{\check{\eta}_{\beta}^{(k+1)} \mid \mu_\beta^{(k+1)} }  \le  c_1 \,,\, \Ent\p{ \mu_{\beta}^{(k+1)} \mid \mu_{\beta^{(k)}}^{(k)} } \le  c_2\,,\,\beta^{(k)} \le \beta \le \beta_\infty  \}.
 \end{equation}
However the feasible set $\mathcal{C}(k)$ might be an empty set. The idea is thus to extend  problem \eqref{eq:bridgeProblem0} to indices $k' \in \llbracket 0, k \rrbracket$, and  search for the largest index $k'$ such that  the set $\mathcal{C}(k')\neq \emptyset$, by keeping in memory  the past proposal distributions $\{\mu_{\beta^{(k')}}^{(k')}\}_{k'\le k}$.
In order to set a feasible bridging  inverse temperature $\beta_{\mrm{bridge}}(k+1)$, it  is  then natural to consider the problem 
\begin{align}\label{eq:bridgeProb}
\text{Find: } \beta \in  \mathcal{C}(k_{\mrm{bridge}})\quad\text{where}\quad k_{\mrm{bridge}} \eqdef  
\textrm{largest} \, k': \, k' \in \llbracket 0, k \rrbracket \text{ s.t.  }  \mathcal{C}(k')\neq \emptyset.
\end{align} 
In practice, the set $\mathcal{C}(k_{\mrm{bridge}})$ is not a singleton and we  need to define $\beta_{\mrm{bridge}}(k+1)$ as a particular element of this set. One possible choice for $\beta_{\mrm{bridge}}(k+1)$, which we will use, is to select the largest element of $\mathcal{C}(k_{\mrm{bridge}})$.   Note that for $k'=0$, the initial proposal distribution  is related to a vanishing inverse temperature $\beta^{(0)}=0$. Therefore $\mathcal{C}(0)=\{0\}$, which shows that a solution of \eqref{eq:bridgeProb} always exists.  

It is straightforward to show that, at the bridging temperature solution $\beta_{\mrm{bridge}}(k+1)$, the normalization constant $Z^{(k+1)}_{\beta_{\mrm{bridge}}(k+1)}$ related to  the   distribution $\mu^{(k+1)}_{\beta_{\mrm{bridge}}(k+1)}$ satisfies 
\begin{equation}\label{eq:bridgeZ}
Z^{(k+1)}_{\beta_{\mrm{bridge}}(k+1)}=Z^{(k_{\mrm{bridge}})}_{\beta^{(k_{\mrm{bridge}})}}\mu^{(k_{\mrm{bridge}})}_{\beta^{(k_{\mrm{bridge}})}}\left(e^{{\beta_{\mrm{bridge}}(k+1)}\Srom{(k+1)}(x)-\beta^{(k_{\mrm{bridge}})}\Srom{(k_{\mrm{bridge}})}(x)}\right).
\end{equation}
The adaptive tempering procedure discussed in Section~\ref{sec:adaptempering}  can then be deployed:  the  recursion is initialized  at iteration $k+1$ using the reduced score  for  $S^{(k+1)}$ as  $\beta_0:=\beta_{\mrm{bridge}}(k+1)$ (see definition \eqref{eq:ATcrietrion}), and is run up to the critical parameter $\beta_n:= \beta^{(k+1)}(c_1,\beta_{\mrm{bridge}}(k+1) )$  (as given by definition \eqref{eq:level}).

\subsection{Stopping criterion}\label{sec:stop}
In principle, the algorithm can continue as long as one wishes, for instance until the budget of snapshot samples is exhausted. 
 However, in practice, we stop updating the reduced score of the algorithm if the divergence between the surrogate target and the reduced proposal is very small and if the associated critical inverse temperature equals $\beta_\infty$.  More precisely, 
 no update of the reduced model is performed if the following condition is satisfied:
\begin{equation}\label{eq:stopUpdate}
 \Ent\p{\check{\eta}_{\beta^{(k)}}^{(k)} \mid \mu_{\beta^{(k)}}^{(k)} }  \le \epsilon \quad \& \quad \beta^{(k)}= \beta_\infty,
\end{equation}
where  $\epsilon>0$ is a given threshold parameter close to the machine precision. In that case the algorithm amounts to a standard importance sampling procedure based on the proposal $\mu_{\beta_\infty}$ computed with the same, final, reduced model.

%

\subsection{The  algorithm}\label{subsec:pseudo}
The algorithm presented below gathers the ingredients presented in the previous sections.  It is an idealized pseudo-code and,  in practice, it is not directly implementable. Indeed, it assumes exact evaluation of expectations and exact sampling according to the proposals. In Section~\ref{sec:practical}, we will propose a practical implementation of this idealized algorithm using a sample of $N$ clones (or \tit{particles}) in SMC methodology. The latter will simulate the flow of proposals for diminishing  temperatures using  Monte Carlo samples with approximately correct distributions for large sample sizes $N \to + \infty$. \medskip

As shown in Appendix~\ref{app:martingal} with a martingale argument, the unnormalized estimator~\eqref{eq:unestimator} computed by the  algorithm below is unbiased
$$
\E \, \hat{\gamma}_{\beta_\infty}^{H}  = \gamma_{\beta_\infty}^\ast,
$$
and its variance scales in $\bigO(H^{-1})$. Note that in this idealized setup, samples  are assumed to be  drawn exactly from the sequence of importance distributions. In practice,  as detailed in  Section~\ref{sec:practical}, this assumption will not hold exactly because of the adaptive features used in the sampling of the  proposals by the SMC routine, yielding a small bias of order the inverse of the sample size.\medskip

{\bf Algorithm (ART)}
{\small 
\begin{algorithmic}[1]
 \Require  inverse temperature target $\beta_\infty$, initial reduced score $\Srom^{(1)}$ and error $\Err^{(1)}$, budget of snapshots $K$, worst-case log cost threshold $c_1$, adaptive tempering parameter $c_2$, reduced score precision $\epsilon \ll c_1$, trigger parameter $j_0$, learning intensity $\tau$.

\Comment{{(\color{blue} initialization)}}

\State Set  $\beta_{\mrm{bridge}}(1)=0$, $H=0$, $j=0$, $k=0$, ${Z_\beta}^{(1)}=1$ 
\While{$K>0$}
\State{$k \leftarrow k+1$}

\Comment{{(\color{blue} critical temperature and adaptive tempering, see Section~\ref{sec:prop_temp}~\&~\ref{sec:adaptempering})}}
\State Simulate the flow of proposals $\mu^{(k)}_{\beta_i}$ for $i=0,\ldots,n$ using the recursion 
\begin{align}\label{eq:algoSMC}
\left\{
\begin{array}{ll}
 \beta_0=\beta_{\mrm{bridge}}(k)\\
 \beta_{i+1}= \min \{  \beta : \; \Ent\p{ \mu_{\beta}^{(k)} \mid \mu_{\beta_{i}}^{(k)} } >  c_2 \,,\,  \beta_i \leq \beta \le \beta_\infty\},\\
  \end{array}
\right.
\end{align}
where the index of the final inverse temperature $ \beta_n$  is 
\begin{align}\label{eq:algoCT}
n=\min \{ i : \Ent\p{\check{\eta}_{\beta_{i+1}}^{(k)} \mid \mu_{\beta_{i+1}}^{(k)} }  >  c_1\quad \textrm{or}\quad \beta_i = \beta_\infty\}.
\end{align}
 \State Set $\beta^{(k)}=\beta_n$ and compute  the associated normalisation $Z^{(k)}_{\beta^{(k)}}$ with \eqref{normalizationRec}.

%
%

\Comment{{\color{blue} (snapshot sampling, see Section~\ref{sec:learn})}}
 \State {\bf if} {$\beta^{(k)}=\beta_\infty $}  {\bf then}  $j \leftarrow j+1$  {\bf end if}
\State  {\bf if}   $j\ge j_0$ {\bf then} draw 
$ X \sim \mu_{\beta^{(k)}}^{(k)}, $ {\bf else}  $X \sim \e^{\tau E^{(k)}} \mu_{\beta^{(k)}}^{(k)}/Z^{(k)}_\tau$ {\bf end if}
\State Compute  the snapshot  $\Score(X)=\mrm{score}( \Psi(X))$

\Comment{{\color{blue} (stopping criterion, see Section~\ref{sec:stop})}}

 \State{\bf if}{ $ \Ent(  \check \eta^{(k)}_{\beta^{(k)}} \mid  \mu^{(k)}_{\beta^{(k)}} ) >\epsilon$ or  $\beta^{(k)} < \beta_\infty$}{\bf then} \\
 \hspace{0.5cm}Update  reduced model: {build $(\Srom^{(k+1)} ,\Err^{(k+1)} )$ by enriching  $(\Srom^{(k)} ,\Err^{(k)} )$ with   $(X,\Score(X))$}\\
  \hspace{0.4cm} {\bf else} Don't update the reduced model: set $(\Srom^{(k+1)} ,\Err^{(k+1)} ) \leftarrow (\Srom^{(k)} ,\Err^{(k)} )$  {\bf end if}
 
\Comment{{(\color{blue} importance sampling, see Section~\ref{sec:IS})}}
\If{ $j>j_0$}
\State{$H\leftarrow H+1$}
\State {$$ \hat \gamma^H_{\beta_\infty} \leftarrow \frac{(H-1)}{H}\hat \gamma^{H-1}_{\beta_\infty} + \frac{1}{H}Z^{(k)}_{\beta^{(k)}}e^{\beta_\infty\Score(X) - \beta^{(k)}\Srom^{(k)}(X)} \delta_X,$$}
  \EndIf
  \State Set $K\leftarrow K-1$
 
 \Comment{{(\color{blue} bridging, see Section~\ref{sec:bridging})}}
 \State Find the largest $k' \in \llbracket 0, k \rrbracket$ such that exists $\beta' \in \mathcal{C}(k')$ where: 
$$
\mathcal{C}(k') = \{ \beta'  : \; \Ent\p{\check{\eta}_{\beta'}^{(k+1)} \mid \mu_{\beta'}^{(k+1)} }  \le  c_1 \,,\, \Ent\p{ \mu_{\beta'}^{(k+1)} \mid \mu_{\beta^{(k')}}^{(k')} } \le  c_2\,,\,\beta^{(k')} \le \beta' \le \beta_\infty  \}
$$
  and denote a solution pair by 
 $
 (k_{\mrm{bridge}}(k+1),\beta_{\mrm{bridge}}(k+1) ).
 $

 \State {Simulate the proposal $\mu^{(k+1)}_{\beta_{\mrm{bridge}}(k+1)}$ starting from $\mu^{(k_{\mrm{bridge}}(k+1))}_{\beta^{k_{\mrm{bridge}}(k+1)}}$}
 \State{Compute using \eqref{eq:bridgeZ} the normalization $Z^{(k+1)}_{\beta_{\mrm{bridge}}(k+1)}$}. 
%
%

\EndWhile

\State \textbf{Return}   $\hat \gamma^{H}_{\beta_\infty} / \hat \gamma^{H}_{\beta_\infty}(\one) $
\end{algorithmic}
}\medskip

\section{Algorithm in Practice: SMC Simulation}\label{sec:practical}
We describe now how an iteration of the idealized ART algorithm presented above may be implemented  in practice.  Suppose we start  the $k$-th iteration of the ART algorithm  with a given sample of size $N$ approximately distributed  according to the law $\mu^{(k)}_{\beta_{\mrm{bridge}}(k)}$, as well as with a given estimate of the associated normalization $Z^{(k)}_{\beta_{\mrm{bridge}}(k)}$.  
We will use in practice a SMC sampling methodology~\cite{Beskos2016} to approximate  the adaptive tempering step introduced in Section~\ref{sec:adaptempering} and the bridging procedure proposed in Section~\eqref{sec:bridging}. The following sections describe these approximations, with the other steps in the algorithm approximated using SMC flowing directly from them. 

\subsection{Approximation of a flow of tempered distribution}\label{sec:AMS}

The SMC simulation generates a finite sequence of $N$-samples approximately distributed according to a sequence of  distributions $\{\mu^{(k)}_{\beta_i}\}_{ 1 \leq i \leq n}$, as well as estimators of the related normalizations $ \set{ Z^{(k)}_{\beta_i} }_{ 1 \leq i \leq n}$. The $i$-th  empirical distribution related to the  $N$-sample $(\Xi^{(i)}_{1}, \ldots, \Xi^{(i)}_{N})$  (we omit in notation the dependence to the current iteration $k$) generated by SMC will be denoted 
$$
 {\mu^{N,{(k)}}_{\beta_i}} \eqdef \frac1N \sum_{\ell=1}^N \delta_{\Xi^{(i)}_{\ell}} \simeq  \mu_{\beta_i}^{(k)}, 
 $$
and the estimator of the normalization constant  $ Z^{(k)}_{\beta_n} $  will be denoted $ {Z^{N,{(k)}}_{\beta_n}}$. \medskip

 The sequence of inverse temperatures  $\beta_0=\beta_{\mrm{bridge}}(k)<\beta_1<\ldots < \beta_n = \beta^{(k)}$ will be redefined in this SMC simulation framework  as an approximation targeting  the  sequence defined in \eqref{eq:algoSMC}, with the final element being an approximation of  the critical   inverse temperature~\eqref{eq:algoCT}.
However, before going into the latter approximations, let us begin by describing how starting from $ {\mu^{N,{(k)}}_{\beta_{i}}} $,   SMC simulates  at the $i$-th iteration the empirical distribution $ {\mu^{N,{(k)}}_{\beta_{i+1}}} $. SMC performs successively:  {\it i)} a weighting and selection step of the $N$ particles, and {\it ii)} a mutation step of each individual particle. The weighting step {\it i)} assigns weights to the empirical distribution ${\mu^{N,{(k)}}_{\beta_i}}$ so that it approximates $\mu_{\beta_{i+1}}^{(k)}$
$$ {\mu_{\beta_{i+1}}^{N,(k)}} = \sum_{\ell=1}^N w^{(i)}_\ell \delta_{\Xi^{(i)}_{\ell}} \quad\textrm{with}\quad  w_\ell^{(i)}=\frac{G_i({\Xi^{(i)}_{\ell}})}{\sum_{\ell'=1}^N G_i(({\Xi^{(i)}_{\ell'}})) },$$ 
where $G_i(x)$ is given by \eqref{normalizationRec}. 
It is straightforward to show that the associated normalizations follow the recursion
$$ {Z^{N,{(k)}}_{\beta_{i+1}} }  = {Z^{N,{(k)}}_{\beta_{i}}}{\mu_{\beta_{i}}^{N,(k)}}(G_i).$$ 
The above weighting step is followed by a selection step, where the particles are resampled according to their weights $w_\ell^{(i)}$. For this purpose, we choose the efficient systematic resampling scheme among the  multiple available resampling techniques~\cite{douc2005comparison}.  The mutation step {\it ii)} then consists of a Markovian transition that leaves $\mu_{\beta_{i+1}}^{(k)}$ invariant and is applied individually and independently a certain number of times to the $N$ particles. In practice, the Markovian transitions are  performed using a Metropolis-Hastings algorithm, with an adaptive Markovian kernel  in order to maintain a given Metropolis acceptance rate~\cite{garthwaite2016adaptive}.   

We emphasize that  tempering with SMC typically involves many evaluations of the score function underlying the  sequence of targeted  distributions. This paper precisely focuses on cases where the true score $\Score$ is too expensive for so many evaluations and is substituted by  a reduced score $\Srom^{(k)}$, much cheaper to evaluate.

\subsection{Approximation of critical temperature and adaptive tempering}\label{sec:}
The simulation starts from an i.i.d. sample with the reference distribution ($\beta_{\mrm{bridge}}(1)=0$ so that $\mu_{\beta_{\mrm{bridge}}(1)}^{(1)}=\pi$).
 We recall that after $k$ iterations, the algorithm provides a current reduced score $\Srom^{(k)}$, an associated point-wise error estimate $\Err^{(k)}$, an initial inverse temperature $\beta_{\mrm{bridge}}(k) $  and a sample of size $N$ approximately distributed according to $\mu_{\beta_{\mrm{bridge}}(k)}^{(k)}$,  denoted using a sample-size superscript by $\mu_{\beta_{\mrm{bridge}}(k)}^{N,(k)}$. \medskip
 
 The ART algorithm needs in practice an approximation of  the critical inverse temperature $\beta^{(k)}$ given by  the final inverse temperature of index determined by the criterion \eqref{eq:algoCT}.  We approximate  this  criterion by substituting the continuous distributions  by sample-based empirical approximations. More precisely, at the $i$-th iteration of the tempering scheme, we  use the empirical distribution  ${\mu^{N,{(k)}}_{\beta_i}}$ to approximate the worst-case log cost for an inverse temperature $\beta$  close to $\beta_i$ as
 \begin{align}\label{eq:criticTempApprox}
 \Ent\p{\check{\eta}_{\beta}^{(k)} \mid \mu_{\beta}^{(k)} }&= \ln\frac{ Z_\beta^{(k)}}{\check Z_\beta^{(k)}} -\beta \check{\eta}_{\beta}^{(k)} (\Err^{(k)}),\nonumber \\
 &\approx  \ln \frac{\mu^{N,{(k)}}_{\beta_i}(e^{(\beta-\beta_i)\Srom^{(k)}})}{\mu^{N,{(k)}}_{\beta_i}(e^{(\beta-\beta_i)\Srom^{(k)}-\beta\Err^{(k)}})} 
 -\beta \frac{\mu^{N,{(k)}}_{\beta_i}(e^{(\beta-\beta_i)\Srom^{(k)}-\beta\Err^{(k)}}\Err^{(k)})}{\mu^{N,{(k)}}_{\beta_i}(e^{(\beta-\beta_i)\Srom^{(k)}-\beta\Err^{(k)}})},
\end{align}
where we have used the definitions of the proposal \eqref{eq:propa} and the surrogate target \eqref{eq:pessTarget}.

The ART algorithm also needs in practice a Monte-Carlo approximation of the  sequence of temperature given in ~\eqref{eq:algoSMC}. Analogously to what was done previously,  at the $i$-th iteration of the tempering scheme, we  use the empirical distribution  ${\mu^{N,{(k)}}_{\beta_i}}$ to approximate the adaptive criterion for $\beta$  close to $\beta_i$ as
\begin{align}\label{eq:tempApprox}
 \Ent\p{\mu_{\beta}^{(k)} \mid \mu_{\beta_{i}}^{(k)} }&= \ln\frac{ Z_{\beta_i}^{(k)}}{ Z_\beta^{(k)}} +(\beta-\beta_i) {\mu}_{\beta}^{(k)} (\Srom^{(k)}), \nonumber \\
 &\approx  -\ln {\mu^{N,{(k)}}_{\beta_i}(e^{(\beta-\beta_i)\Srom^{(k)}})}{} 
 +(\beta-\beta_i) \frac{\mu^{N,{(k)}}_{\beta_i}(e^{(\beta-\beta_i)\Srom^{(k)}}\Srom^{(k)})}{\mu^{N,{(k)}}_{\beta_i}(e^{(\beta-\beta_i)\Srom^{(k)}})},
\end{align}

Now, using these two approximations, the SMC simulation proceeds as explained in Section~\ref{sec:AMS}: it creates a random sequence $(\beta_i)_{i\geq 0}$ of increasing  inverse temperature starting from $\beta_0={\beta_{\mrm{bridge}}(k)}$, related to  a sequence of empirical distributions ${\mu^{N,{(k)}}_{\beta_i}}$ for $i \geq 0$, until it reaches the critical inverse temperature $\beta^{(k)}(c_1,{\beta_{\mrm{bridge}}(k)})$. More precisely, at the $i$-th iteration of the tempering scheme, we propose an increase for $\beta_i$ by $\delta\beta$ given as  the solution of an optimization  problem  defined with the approximation \eqref{eq:tempApprox} 
$$
 \delta \beta=\min \{\delta \beta' :  \delta \beta' \frac{\mu^{N,{(k)}}_{\beta_i}(e^{\delta \beta'\Srom^{(k)}}\Srom^{(k)})}{\mu^{N,{(k)}}_{\beta_i}(e^{\delta \beta'\Srom^{(k)}})}-\ln {\mu^{N,{(k)}}_{\beta_i}(e^{\delta \beta'\Srom^{(k)}})} > c_2,  0\le \delta \beta'\le \beta_\infty-\beta_i\}.$$
This  one-dimensional  problem can be efficiently solved using a standard bisection method. The increase $ \delta \beta$ is accepted  if it is below or equal the critical inverse temperature, \ie using approximation \eqref{eq:criticTempApprox} if 
\begin{align}\label{eq:smcCritique}
\ln \frac{\mu^{N,{(k)}}_{\beta_i}(e^{\delta \beta\Srom^{(k)}})}{\mu^{N,{(k)}}_{\beta_i}(e^{\delta\beta\Srom^{(k)}-(\beta_i+\delta\beta)\Err^{(k)}})} 
- (\beta_i+\delta\beta) \frac{\mu^{N,{(k)}}_{\beta_i}(e^{\delta \beta\Srom^{(k)}-( \beta_i+\delta\beta)\Err^{(k)}}\Err^{(k)})}{\mu^{N,{(k)}}_{\beta_i}(e^{\delta\beta\Srom^{(k)}-( \beta_i+\delta\beta)\Err^{(k)}})}\le c_1. 
\end{align}
Therefore if \eqref{eq:smcCritique} holds then
 $\beta_{i+1}=\beta_i+\delta\beta$, else  the critical inverse temperature is reached, \ie $\beta^{(k)}(c_1,{\beta_{\mrm{bridge}}(k)})=\beta_i$.

\subsection{Approximation for bridging}\label{sec:bridge}


As exposed in Section~\ref{sec:bridging}, the bridging procedure at the end of iteration $k$ consists in selecting a past index $=k_{\mrm{bridge}}\le k$ and a new initial inverse temperature  ${\beta_{\mrm{bridge}}(k+1)}$ which will be used to initialize the SMC simulation with the new score approximation $\Srom^{(k+1)}$ starting from the past, recorded, empirical distribution $\mu^{N,{(k_{\mrm{bridge}})}}_{\beta^{(k_{\mrm{bridge}})}}$.  In line with the definition of problem~\eqref{eq:bridgeProb},  we select for this purpose  the largest lexicographic  pair $(k',\beta)  \in \llbracket 0, k \rrbracket \times \mathcal{C}(k') $. However, in order to provide an implementation of this step of the ART algorithm, it is  necessary to define an empirical counterpart to this optimization problem. 
As done previously, using $\mu^{N,{(k')}}_{\beta^{(k')}}$, the worst log-cost is approximated   for $\beta$ close to ${\beta^{(k')}}$  by 
\begin{align}\label{eq:criticTempApproxBridge}
 \Ent\p{\check{\eta}_{\beta}^{(k+1)} \mid \mu_{\beta}^{(k+1)} }&\approx  \ln \frac{\mu^{N,{(k')}}_{\beta^{(k')}}(e^{\phi_{\beta,k'}})}{\mu^{N,{(k')}}_{\beta^{(k')}}(e^{\phi_{\beta,k'}-\beta\Err^{(k+1)}})} -\beta \frac{\mu^{N,{(k')}}_{\beta^{(k')}}(e^{\phi_{\beta,k'}-\beta\Err^{(k+1)}}\Err^{(k+1)})}{\mu^{N,{(k')}}_{\beta^{(k')}}(e^{\phi_{\beta,k'}-\beta\Err^{(k+1)}})},
\end{align}
while  the entropic adaptive criterion  is approximated  for $\beta$ close to ${\beta^{(k')}}$   by 
\begin{align}\label{eq:tempApproxBridge}
 \Ent\p{\mu_{\beta}^{(k+1)} \mid \mu_{\beta^{(k')}}^{(k')} }\approx  -\ln {\mu^{N,{(k')}}_{\beta^{(k')}}(e^{\phi_{\beta,k'}})}
 + \frac{\mu^{N,{(k')}}_{\beta^{(k')}}(e^{\phi_{\beta,k'}}{\phi_{\beta,k'}})}{\mu^{N,{(k')}}_{\beta^{(k')}}(e^{\phi_{\beta,k'}})},
\end{align}
with $\phi_{\beta,k'}={{\beta\Srom^{(k+1)}-\beta^{(k')}\Srom^{(k')}}}.$
A SMC approximation of the bridging procedure  \eqref{eq:bridgeProb}  consists then in solving a sequence of one-dimensional constrained  optimization problem of the form
$$\beta'=\max\{\beta : \eqref{eq:criticTempApproxBridge} \le c_1, \eqref{eq:tempApproxBridge} \le c_2 , \beta^{(k')} \le \beta \le \beta_\infty\},$$
 starting from index $k'=k$ and decreasing the index until the problem above admits a solution, in which case  $$(k_{\mrm{bridge}},{\beta_{\mrm{bridge}}(k+1)})=(k',\beta').$$
 We remark  that  the set of constraints related to this  problem is not necessarily convex. However, we can efficiently solve this one-dimensional problem with standard optimization techniques. We choose to use an interior point method.
Analogously to the procedure detailed in Section~\ref{sec:AMS}, the  normalization estimate is then  updated as
\begin{equation}\label{eq:inclusionZRefined}
{Z^{N,{(k+1)}}_{\beta_{\mrm{bridge}}(k+1)}}= {Z^{N,{({ k_{\mrm{bridge}}})}}_{\beta^{( k_{\mrm{bridge}})}}} {\mu_{\beta^{( k_{\mrm{bridge}})}}^{N,{({ k_{\mrm{bridge}}})}}}(G_{\mrm{bridge}}),
\end{equation}
where 
$$G_{\mrm{bridge}}(x)=e^{{\beta_{\mrm{bridge}}(k+1)}\Srom^{(k+1)}(x)-{\beta^{( k_{\mrm{bridge}})}\Srom{(k_{\mrm{bridge}})}(x)}}.$$
The     $N$-sample $(\Xi_{1}, \ldots, \Xi_{N})$  related to the empirical distribution ${\mu_{\beta^{( k_{\mrm{bridge}})}}^{N,{({ k_{\mrm{bridge}}})}}}$ is  resampled according to the weights 
$\frac{G_{\mrm{bridge}}({\Xi_{i}})}{\sum_{i'=1}^N G_{\mrm{bridge}}(({\Xi_{i'}})) },$ 
and  mutated to  target the new  initial distribution  $ \mu_{\beta_{\mrm{bridge}}(k+1)}^{(k+1)}$ serving at the next  iteration.\medskip

We omit to comment the SMC approximation for snapshot sampling or computing the stopping index in the ART algorithm as they follow naturally from the above discussion or from the simple substitution of distributions by their empirical equivalents.
This concludes the $k$-th iteration of the algorithm.

\subsection{SMC-based estimators}

 Using the sequence of empirical distributions $\{ \mu_{{\beta^{(k)}}}^{{N,(k)}}\}_{k=1}^K$ generated by SMC simulations along the ART algorithm iterations, we can write a closed-form approximation of the importance sampling estimator given in~\eqref{eq:unestimator}~-~\eqref{eq:estimator}.  The SMC approximation of the IS estimator  $\hat \gamma^{H}_{\beta_\infty}$  of  the  the unnormalized distribution $ \gamma^\ast_{\beta_\infty}$   is given  by
 \begin{align}\label{eq:unestimatorPractice}
  \hat \gamma^{N,H}_{\beta_\infty,IS}=\frac{1}{H}\sum_{k=K_{j_0}+1}^{K_{j_0}+H} Z^{N,(k)}_{\beta^{(k)}}e^{\beta_\infty\Score(X_{k+1}) - \beta^{(k)}\Srom^{(k)}(X_{k+1})} \delta_{X_{k+1}}.
    \end{align}
Moreover, SMC simulation  also provides as by-products an estimator of $ \gamma^\ast_{\beta_\infty}$, which we will call reduced SMC (RSMC) estimator because it is based solely on the reduced score, given by 
\begin{equation}\label{eq:AMSpractice}
\hat \gamma^{N,H}_{\beta_\infty,RSMC} =
\frac{1}{H\,N}\sum_{k=K_{j_0}+1}^{K_{j_0}+H}  {Z_{\beta^{(k)}}^{N,(k)}}\sum_{i=1}^N e^{(\beta_\infty-\beta^{(k)})\Srom^{(k)}}  \delta_{\Xi^{(k)}_i} ,
\end{equation}
where $(\Xi^{(k)}_1,\ldots, \Xi^{(k)}_N)$ denotes the $N$-sample related to the empirical distribution $\mu_{\beta^{(k)}}^{N,(k)}$.
 $\hat \gamma^{N,H}_{\beta_\infty,RSMC}$ is expected to be  increasingly accurate  in the case where the reduced score $\Srom^{(k)}$ converges to $\Score$ as $k$ increases. However, we will see in our numerical experiments that the RSMC estimator can nevertheless retain a significant bias in some cases, even if the reduced score converges to the true one.  The  SMC-based estimates of the normalized distribution $\eta^\ast_{\beta_\infty}$ are  
\begin{equation}\label{eq:normEstimatesSMC}
 \hat \eta^{N,H}_{\beta_\infty,IS} = \hat \gamma^{N,H}_{\beta_\infty,IS}/  \hat \gamma^{N,H}_{\beta_\infty,IS}(\one) \quad\textrm{and}\quad\hat \eta^{N,H}_{\beta_\infty,RSMC}=\hat\gamma^{N,H}_{\beta_\infty,RSMC} / \hat\gamma^{N,H}_{\beta_\infty,RSMC}(\one). 
 \end{equation}

\section{Numerical Simulations}\label{sec:num}

Our numerical simulations  focus on target distributions of the form \eqref{eq:target}, where $\Score$ implies  the high-dimensional solution of an elliptic  PDE  parametrized by  some random input vector.  We propose to evaluate numerically the ART algorithm for the sampling of these distributions, where  the PDE solution is approximated by a popular  reduced basis (RB) approach. The evaluation is completed by a study of the ART's performance in the context of a one-dimensional yet challenging toy problem. Our numerical simulations are carried out using an implementation of the ART algorithm in the Julia language, available at \url{https://gitlab.inria.fr/pheas/adaptive-reduced-tempering}. After introducing the models, namely the PDE and its high-fidelity and reduced solutions, we will study two  related  problems: a rare event estimation problem and  a Bayesian inverse problem.

\subsection{Model \#RB: heat diffusion and reduced basis approximations}\label{sec:PDE}

\begin{figure}[h!]
\begin{center}
\begin{tabular}{cc}
\includegraphics[width=0.35\textwidth]{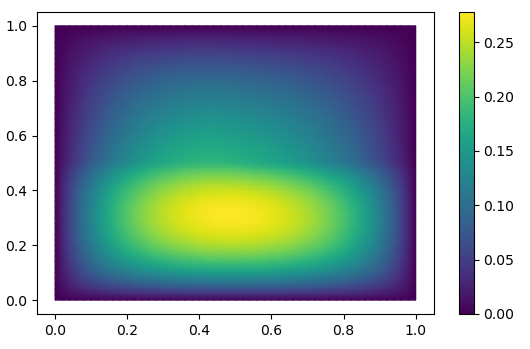}&\includegraphics[width=0.35\textwidth]{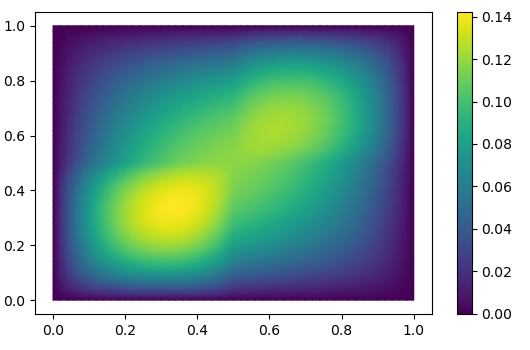}\\
\includegraphics[width=0.35\textwidth]{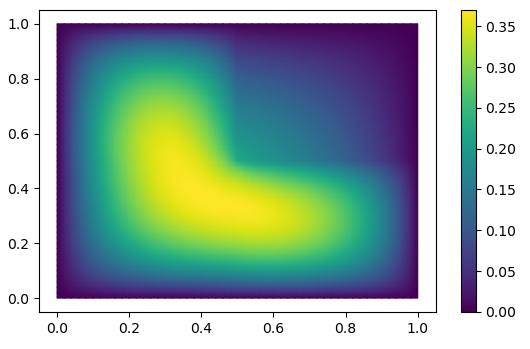}&\includegraphics[width=0.35\textwidth]{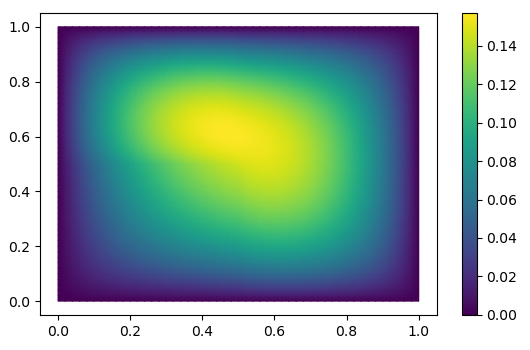}\vspace{-0.35cm}
\end{tabular}
	\caption{{\footnotesize {\bf Illustration of the thermal block problem (Model \#RB).}   High-fidelity solution $\Psi^\ast(X)$ of the PDE with $d=4$ for $4$ different samples $X$ of the reference log-normal distribution (images taken from~\cite{cerou2024adaptive}).
	 {  \label{fig:4}}}}
		\end{center}\vspace{-0.cm}
\end{figure}

Let  $\Psi^\ast(x)$ be   the high-fidelity numerical approximation of the solution of a  bi-dimensional PDE on the spatial domain  $\Omega=[0,1]^2$ parametrized by  a  $d$-dimensional random vector $x=(x_1\cdots x_d)^\intercal\in  \mathcal{X}=\R_+^d$. We will consider $d=2\times2$ or $d=5\times 5$ in our experiments. The chosen PDE corresponds to the well-studied  {\it thermal block problem}, which models heat diffusion  in a heterogeneous medium.

We begin by describing the parametric PDE and the high-fidelity numerical approximation of its solution. The temperature function $\Psi$ over  $\Omega$ is ruled by the elliptic problem
\begin{equation*}
 \left\{\begin{aligned}
-\nabla \cdot(\kappa(\cdot;x) \nabla \Psi)=&1\quad  \textrm{in}\quad \Omega\\
\Psi=&0\quad  \textrm{on}\quad \partial \Omega
\end{aligned}\right.,
\end{equation*}
where the function $\kappa (\cdot;x):\Omega \to \mathbb{R}_+$ represents a parametric diffusion coefficient function.

The   high-fidelity solution $\Psi^\ast(x)$  of this diffusion problem entails the solution of a variational problem.  Precisely, let $\mathcal{V}$ denote the Hilbert space $\mathcal{H}^1_0(\Omega)$ and $\|\nu\|_\mathcal{V}= \| \nabla \nu \|_{L^2}$ the norm induced by the inner product defined over $\mathcal{V}$. For any $x \in \mathcal{X}$, we define the solution $\Psi^\star(x)\in \mathcal{V}$  of the variational problem
$a(\Psi(x),\nu;x)=\int_\Omega  \nu \d\Omega,\quad\forall \nu \in \mathcal{V},$
where  $a: \mathcal{V} \times \mathcal{V} \to \mathbb{R}$ is the bilinear form
$a(u,\nu;x)=\int_\Omega \kappa(\cdot;x) \nabla u \cdot \nabla \nu d\Omega$. We consider the piecewise constant  diffusion coefficient $\kappa(\cdot;x)=\sum_{q=1}^{d} \one_{ \Omega_q}(\cdot) x_q$ with the partition $\Omega=\cup_{q=1}^{d}\Omega_q$ and $\Omega_q$'s being non-overlapping squares. 
Setting the parameter space $\mathcal{X}= (\mathbb{R}_+)^d$ implies that   the  bilinear form is  strongly coercive, and the  Lax-Milgram lemma  ensures the variational problem has a unique solution.

 Consider next a finite $h$-dimensional subspace $\mathcal{V}_h \subset \mathcal{V}$. We will set $h=5101$ in our experiments. The  high-fidelity approximation $\Psi^\ast(x)\in \mathcal{V}_h$ of the solution $\Psi^\star(x)$  is defined for any $x\in \mathcal{X}$ by 
\begin{align*}
\Psi^\ast(x)=\arg\min_{\nu \in \mathcal{V}_h} \| \Psi^\star(x)-\nu\|^2,
\end{align*}
with the energy norm induced by the bilinear form $\| \nu \|=\sqrt{a(\nu,\nu;x)}$. The  high-fidelity approximation $\Psi^\ast(x)$ is obtained in practice by solving a  large  system of $h$ linear equations~\cite{quarteroni2015reduced}. 
 Figure~\ref{fig:4} shows several  high-fidelity solutions  $\Psi^\ast(x)$ of the PDE for  samples drawn according to the reference distribution. 

  We are interested in the construction of an inexpensive approximation to the set of high-fidelity solutions  
$$
\mathcal{M}=\{\Psi^\ast(x)\in \mathcal{V}_h : x \in \mathcal{X} \},
$$
by evaluating only some of its elements.  By selecting a well-chosen set of $K$ snapshots $\{\Psi^\ast(x_i)\}_{i=1}^K$, the  RB  framework  approximates $\mathcal{M}$
in a well-chosen linear subspace $ \mathcal{V}_K \subset  \mathcal{V}_h$ of much lower dimension $K \ll h$. 
Using this framework yields    {\it i)} an approximation  $\Psi$ of  the high-fidelity solution $\Psi^\ast$,  {\it ii)} a so-called  {\it a posteriori} error upper bound on the finite element version of the reference $\mathcal{H}^1_0(\Omega)$ norm $\| \Psi^\ast(x)-\Psi(x)\|_{\mathcal{V}_h}$ denoted by
\[
\| \Psi^\ast(x)-\Psi(x)\|_{\mathcal{V}_h} \leq  \Delta_\Psi(x) .
\] 
The high-fidelity solution, the RB approximation and the a posteriori error bounds are computed via  the ``pymor''  Python\textsuperscript{\textregistered} toolbox~\cite{milk2016pymor} available at \url{https://github.com/pymor/pymor}.   

The {\it on-line}  computation of the RB approximation and error bound relies on the affine parametric dependence of the RB approximation and  the {\it off-line}  assembly of the reduced matrices and vectors.  This {\it off-line} phase, which is performed once for all at each RB update, entails the computation of all the $h$-dependent and $x$-independent structures based on the snapshots\footnote{With regard to the snapshots used to enrich the RB approximation space, it should be noted that the sampling of snapshots in the first iterations of ART corresponds to the popular {\it weak greedy} algorithm.   
However, once the target temperature has been reached ($j_0$ times), the weak greedy algorithm is replaced by drawing snapshots according to the importance distribution. 
}. We consider the case of  a Galerkin RB approximation with $K \ll h$ and $d \ll h$. In the case of the finite element discretization of our elliptic problem,  computing scalar products, matrix-vector products or the  high-fidelity solution has a complexity of $O(h)$, {but with a very large constant} (thanks to the {\it generalized minimal residual } method  \cite{Baker05} used to solve sparse linear system in  pymor). It thus requires off-line $\mathcal{O}(hdK^2)$ operations   for computing reduced matrices and vectors used to assemble the RB system, and $\mathcal{O}(hd^2K^2)$ operations for computing those used to build the error estimator. Computing the snapshot in $\mathcal{O}(h)$ operations is in general   less expensive in comparison.  In the latter, on-line phase, for a given parameter $x\in \mathcal{X}$, we solve the RB system  in $\mathcal{O}(K^3+dK^2)$ and provide the error bound  in $\mathcal{O}(d^2K^2)$,  with a cost independent  on $h$.  For further details, we refer the reader to the introductory book on RB~\cite{quarteroni2015reduced}. {We will show in Appendix~\ref{app:ArthighDim} that the dependence in $d^2$ in the complexities for error update and evaluation makes the ART method  non-competitive for $d=25$.}

For  sampling problems involving this PDE model, the  reference distribution $\pi$ for $x$ will be  a log-normal distribution of independent components with $\ln(x_q) \sim \mathcal{N}(0.6,0.8^2)$, for $q=1,\ldots,d$.
 
 \subsection{Model \#S: closed-form score}\label{rem:toy}
From an evaluation perspective, it is difficult to obtain closed-form reference posterior distributions in the context of the previous PDE model. In order to further evaluate the proposed algorithm with a closed-form characterizable reference probability distribution, we rely on the one-dimensional closed-form function proposed in~\cite{cerou2024adaptive}.  In this toy model, the previous high-fidelity PDE solution $\Psi^\ast$  is substituted by a  one-dimensional function $\Psi^\ast: \mathbb{R}_+ \to \mathbb{R}$,  the RB approximation $\Psi$   is replaced by a cubic spline approximation $\Psi: \mathbb{R}_+ \to \mathbb{R}$ and 
we assume that the error $|\Psi(x) - \Psi^\ast(x)|$  is directly available. 
For $x\in\mathcal{X}=\mathbb{R}_+$, the  one-dimensional function is
\begin{equation}\label{eq:1dfunction}
\Psi^\ast(x)=
  \left\{\begin{aligned}
  \ell \quad  &\textrm{if} \quad x \le 1/\ell \\
1/x+f(x) \quad &\textrm{else} 
\end{aligned}\right. ,
\end{equation}
where 
$f(x)=15 \left( \one_{x_0\le x <x_1}  \sin(x-x_0)^2 + \one_{x_1 \le x} ( \sin(x_1-x_0)^2 -0.1(x-x_1))\right)$
with parameters $x_0=0.5$ and $x_1=5$.  

For  sampling problems  involving this toy model, the reference distribution $\pi$ will be    a log-normal distribution such that $\ln(x) \sim \mathcal{N}(1.5,1.5^2)$.

\subsection{Sampling problems}

\begin{figure}[h!]
\begin{center}
\begin{tabular}{ccc}
\hspace{-1cm}{\footnotesize  high-fidelity sol. $\Psi^*$  }&\hspace{-0.5cm}{\footnotesize RB approx. $\Psi$ (initial)  }&\hspace{-0.5cm}{\footnotesize  RB approx. $\Psi$ (final)}\\
\hspace{-1cm}\includegraphics[width=0.32\textwidth]{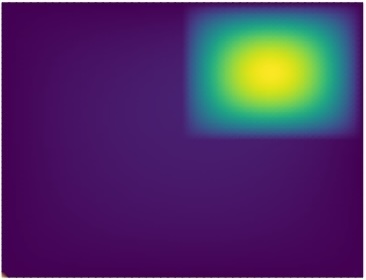}&\includegraphics[width=0.32\textwidth]{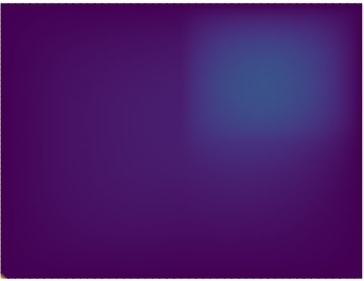}&\includegraphics[width=0.32\textwidth]{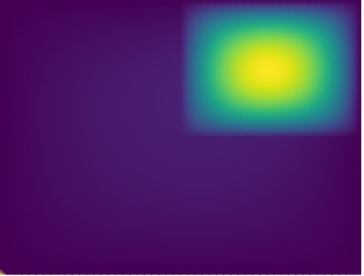}\includegraphics[height=0.22\textwidth]{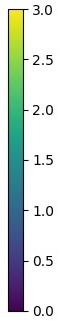}\vspace{-0.cm}\\
\hspace{-1cm}&\hspace{-0.5cm}{\footnotesize Error approx. $\Err_\Psi$ (initial)  }&\hspace{-0.5cm}{\footnotesize  Error approx. $\Err_\Psi$ (final)}\\
\hspace{-1cm}&\includegraphics[width=0.32\textwidth]{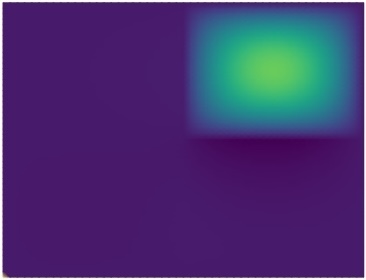}\includegraphics[height=0.22\textwidth]{./ImagesArt/rb/colorbar}&\includegraphics[width=0.32\textwidth]{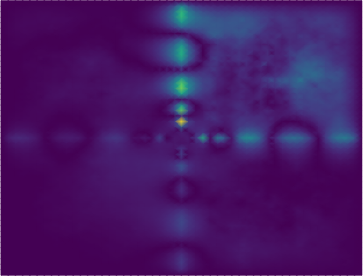}\includegraphics[height=0.22\textwidth]{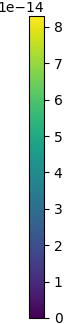}\vspace{-0.35cm}
\end{tabular}
	\caption{{\footnotesize  { {\bf Problems~\#RE with model~\#RB}.   Illustration of the need for an adaptive approach to construct relevant reduced-score approximations of the  PDE solution $\Psi^\ast$.  The RB approximation $\Psi$ and its error $| \Psi-\Psi^\ast|$ are displayed for a sample $X$ in the rare event $\{x: \|\Psi^\ast(x) \|_\infty \ge 3\}$,  at the start and the end of the algorithm (notice the different error scales  in the colorbar).    \label{fig:3}}}}
		\end{center}\vspace{-0.cm}
\end{figure}
\begin{figure}[h!] \label{fig:0}
\begin{center}
\begin{tabular}{cc}
\includegraphics[width=0.45\textwidth]{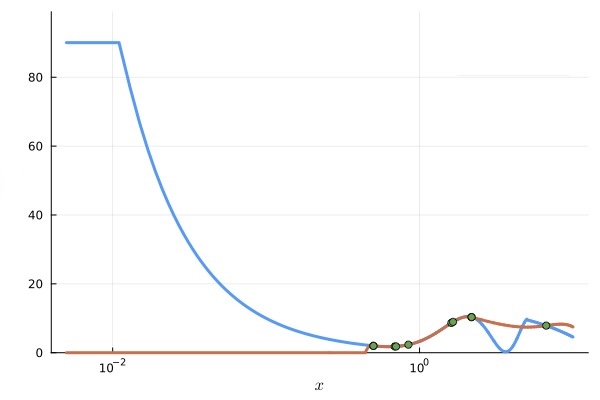}&
\includegraphics[width=0.45\textwidth]{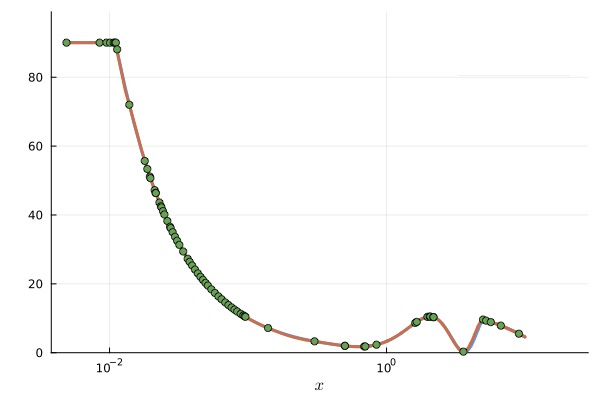}\vspace{-0.5cm}
\end{tabular}
	\caption{{\footnotesize  { {\bf Problem~\#RE with model \#S}.   Illustration of the need for an adaptive approach to construct relevant approximations of $\Psi^\ast$ as defined in  \eqref{eq:1dfunction}.  Function $\Psi^\ast$ (blue curve), its spline approximation $\Psi$ (red curve) and the snapshots used to build it (green dots) are displayed at the start and end of the algorithm. The rare event is  $\{x: \Psi^\ast(x) \ge 90\}$ (the ``plateau'' around $x=1e-2$).    \label{fig:3a}}}}
		\end{center}\vspace{-0.cm}
\end{figure}
 \subsubsection{Problem~\#RE: rare event  simulation}\label{sec:Prob1}
In these numerical experiments, we consider a  target distribution $\eta^\ast_{\beta_\infty}$ of the form \eqref{eq:target}  with 
$\mrm{score}(\cdot)= 1 -\max(\ell-\|\cdot\|_p,0)/\ell$ and $p>0$
yielding the score function 
\begin{align}\label{eq:scoreTheta}
\Score(x) = 1 -\max(\ell-\|\Psi^\ast(x)\|_p,0)/\ell,\quad \ell >0.
\end{align}
The rare event problem is the estimation of the   probability $\lim_{\beta_\infty \to \infty} \pi(e^{\beta_\infty\Score})=\mref\p{ \set{\Score \geq 1}}$.  In practice, we set $\beta_\infty=50$.
 The function $\Psi^\ast(x)$ is  either 
\begin{itemize}
\item  the high-fidelity approximation  of the  PDE solution  for parameter $x$ defined in Section~\ref{sec:PDE} ({\bf model~\#RB}), 
\item the closed-form function \eqref{eq:1dfunction} defined in Section~\ref{rem:toy} ({\bf model~\#S}).
\end{itemize}
Let $p^*=\pi(e^{\beta_\infty\Score})$ denote the target rare event probability. In the case of model~\#RB, this probability corresponds to the occurrence of  reaching the critical temperature $\ell$, \textit{e.g.} a melting temperature, on average  in the domain ($p=1$) or in some point of the domain ($p=\infty$). For this model, we perform standard adaptive SMC simulation with a score based on the high-fidelity solution $\Psi^\ast(x)$ and using $1000$ particles. By averaging over 10 runs,  we  obtain   after an  intensive calculation  empirical  approximations   of the target rare event probabilities: $p^*\approx 6.07e-04$ (for  $d=4$, $p=1$ and $\ell=0.5$), $p^*\approx 1.20e-04$ (for $d=25$, $p=1$ and $\ell=0.2$)  and $p^*\approx 1.47e-04$ (for $d=4$, $p=\infty$ and $\ell=3$). In the case of model \#S,   the target probability $p^*\approx 2.18e-8$ (for $\ell=90$) is closed-form and  analytically  computable.  We use the ART algorithm to estimate  probability $p^*$, evaluating  the unnormalized estimates $\hat \gamma^{N,H}_{\beta_\infty,IS}(\one)$ and $\hat \gamma^{N,H}_{\beta_\infty,RSMC}(\one)$ defined in \eqref{eq:unestimatorPractice} and \eqref{eq:AMSpractice}.

As needed by the ART algorithm, we design the error bound $\Err(x)$ on the absolute error on the score $|\Score(x)-\Srom(x)|$, relying on the available error quantification.  
More precisely, 
  in the case of model~\#S,  as $\Psi(x)\in \mathbb{R}$ and as we have assumed the error is directly available, we simply have that   $ |\Score(x)-\Srom(x)| \le |\Psi(x)-\Psi^\ast(x) |=\Err(x)$.    In the case of model~\#RB with   $p=1$, an error function defined with the a posteriori error $ \Delta_\Psi(x)$ as $\Err(x)=\pi^{-1} \Delta_\Psi(x)$ satisfies 
\begin{align*}
|\Score(x)-\Srom(x)| &\le  |\| \Psi^\ast(x)\|_{L^1}-\|\Psi(x)\|_{L^1}|,\\  &\le   \| \Psi^\ast(x)-\Psi(x)\|_{L^1},\\ &\le   \| \Psi^\ast(x)-\Psi(x)\|_{L^2}  \le \pi^{-1}\| \Psi^\ast(x)-\Psi(x)\|_{\mathcal{V}_h}
 \le \Err(x),
\end{align*}
where we have used the definition of the score function, the triangle inequality, the Cauchy-Schwarz inequality,  the Poincaré inequality with the constante $\pi^{-1}$   on the unit square domain and finally the a posteriori error inequality. As the finite element basis is an interpolating one, we will assume  $h$  sufficiently large so that we can compute the $L^1$-norm as  the  average of the finite element coefficients. In the case of $p=\infty$,  one cannot in general rigorously derive an upper bound on the error relying on the a posteriori estimate  $\Delta_\Psi(x)$. However, as the ART algorithm has the advantage of not requiring a rigorous error bound, we simply define the approximate error bound as 
\begin{align}\label{eq:approxErrorBound}
 |\Score(x)-\Srom(x)|\approx \Delta_\Psi(x)=\Err(x).
\end{align}


Figure~\ref{fig:3} 
 illustrates a high-fidelity approximation  related to the rare event (\ie $\{x:\Score(x)\ge 1\}=\{x:\|\Psi^\ast(x)\|_\infty \ge 3\}$)  obtained by our algorithm in the context of  model~\#RB. We use an initial RB approximation built with $5$ snapshots  drawn according to the reference density. The same figure shows that this initial RB  approximation is insufficiently accurate to discriminate this rare event  since $\{x:\|\Psi(x)\|_\infty \ge \ell\}=\emptyset$. Figure~\ref{fig:3a} shows that similarly, in the case of model~\#S, the initial spline approximation built with 10 snapshots is  insufficiently accurate to discriminate the rare event. 

For these low-temperature problems,  ART performance 
is quantified by evaluating   the relative  expected  square error as a function of the expected cost  over $R$ independent runs. More specifically, let $\hat p_r$   be a short-hand notation for either the estimate $\hat \gamma^{N,H}_{\beta_\infty,IS}(\one)$ or  $\hat \gamma^{N,H}_{\beta_\infty,RSMC}(\one)$ at the $r$-th run.  After exhausting the budget $K$ of snapshots, the relative  expected  square error of the IS or RSMC estimator is
 approximated over runs by the empirical average
\begin{equation}\label{eq:expError}
 \frac{1}{R}\sum_{r=1}^R\frac{( \hat p_{r}-{p^*)}^2}{  {p^*}^2} ,
\end{equation}
 as a function of the expected computational cost  \eqref{eq:expCost} detailed below.

 \subsubsection{Problem~\#BI: Bayesian inverse problem}\label{sec:Prob2}

\begin{figure}[h!]
\begin{center}
\begin{tabular}{cc}
\hspace{-0cm}{\footnotesize  noisy $\Psi^\ast$ }&\hspace{-0.5cm}{\footnotesize random set of observed  points  }\\
\hspace{-0cm}\includegraphics[width=0.4\textwidth]{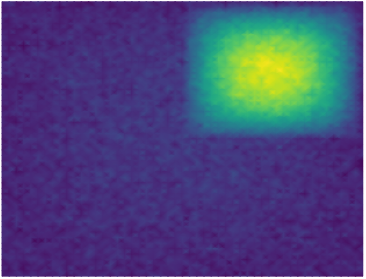}&\includegraphics[width=0.4\textwidth]{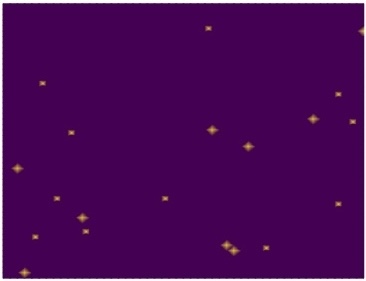}\vspace{-0.35cm}
\end{tabular}
	\caption{{\footnotesize  { {\bf Problem~\#BI with model~\#RB}. Illustration of the observations of the inverse problem. The high-fidelity solution $\Psi^\ast$ for a sample $X$ in the rare event set (obtained at the end of the algorithm for the sampling problem~\#RE) is observed on a random set of points up to Gaussian noise.    \label{fig:d}}}}
		\end{center}\vspace{-0.cm}
\end{figure}

In this numerical experiment, we want to sample a Bayesian posterior distribution  $\eta^\ast_{\beta_\infty}$ of the form \eqref{eq:target} at the inverse temperature $\beta_\infty =1$, for $\pi$ being the  log-normal  prior distribution of model~\#RB,  and given a set of observations following  some  likelihood model. Let  the supremum norm  of the PDE solution and  of its RB approximation evaluated  at the diffusion parameter vector $x$  be denoted  respectively by the functions $\varphi^\ast(\cdot)=\| \Psi^\ast (\cdot) \|_\infty$ and $\varphi(\cdot)=\| \Psi (\cdot) \|_\infty$.  The   objective of this Bayesian sampling problem  is to estimate the posterior distribution of $X$,  given the PDE solution  on a finite set of points corrupted by noise. For evaluation purposes, we will focus in this experiment in the  distribution of the scalar observables $\varphi^\ast(X)$ and $\varphi(X)$.   In order to generate the observations, we consider the high-fidelity PDE solution  $\Psi^{\ast}(x^\star)$ evaluated at  $x^\star$, where  the latter parameter is sampled in the tail of the prior distribution $\pi$ (using a sample of the  low-temperature simulation experiment, see Section~\ref{sec:Prob1}). Let $\Psi^{\ast}_z\in \mathbb{R}$ be the PDE solution $\Psi^{\ast}$ taken at point $z\in \Omega$. Given  a finite set of points $ \mathcal{Z} \subset \Omega $, the set $\{\Psi^{\ast}_z(x^\star)\}_{z\in \mathcal{Z}}$  is corrupted by  an independent Gaussian noise of variance~${\sigma_{obs}^2}$ to generate the set of observations $\{\Psi^{obs}_z\}_{z\in \mathcal{Z}}$.   The  (non-Gaussian) log likelihood of the  observation model is given by the score function 
$$
\Score(x)=-\frac{1}{\sigma_{obs}^2}\sum_{z\in \Omega_{obs}}(\Psi_z^\ast(x)-\Psi^{obs}_z)^2.
$$

We perform standard adaptive SMC simulation with a score based on the high-fidelity solution $\Psi^\ast(x)$ and using $500$ particles. By averaging over $10$ runs,  we  obtain   after an  intensive calculation an empirical  reference approximation   of the   distribution of  $\varphi^\ast(X)$, whose cumulative distribution function will be denoted by $F_{\varphi^\ast}(x)$.  We perform $R$ independent runs of the ART algorithm to estimate the latter function, using  the   normalized distribution estimates  defined in \eqref{eq:normEstimatesSMC}.  More precisely, $F_{\varphi^\ast}(x)$ is approximated using the IS estimate $\hat \eta^{N,H}_{\beta_\infty,IS}$ where the sample $X_{k+1}$ is replaced by $\varphi^\ast(X_{k+1})$, or using the RSMC estimate $\hat \eta^{N,H}_{\beta_\infty,RSMC}$ where the $N$-sample $(\Xi^{(k)}_1,\ldots, \Xi^{(k)}_N)$ is replaced by $( \varphi (\Xi^{(k)}_1) ,\ldots, \varphi (\Xi^{(k)}_N) )$. We will adopt in the following the short-hand notation  $F_{\varphi,r}^{N,H}$ for the cumulative distribution function related to the IS or RSMC estimates of $F_{\varphi^\ast}(x)$, obtained at the  $r$-th run.

 In this experiment, we rely on the reduced score $\Srom(x)=-\sum_{z\in \Omega_{obs}}(\Psi_z(x)-\Psi^{obs}_z)^2/{\sigma_{obs}^2}$ where $\Psi_z(x)$ is the value of the RB approximation $\Psi(x)$ at the point $z\in \Omega$. Using  \eqref{eq:approxErrorBound}, an approximate upper bound is    
$$
|\Score(x)-\Srom(x)| =\frac{1}{{\sigma_{obs}^2}} | \sum_{z\in \Omega_{obs}}(\Psi_z(x)-\Psi_z^\ast(x))(\Psi_z(x)+\Psi_z^\ast(x)-2\Psi^{obs}_z)|\approx  \Err(x),
$$
with    $$ \Err(x)=\frac{1}{{\sigma_{obs}^2}}\sum_{z\in \Omega_{obs}} \Delta_\Psi(x)( \Delta_\Psi(x)+2|\Psi_z(x)-\Psi^{obs}_z|).$$

In this Bayesian sampling experiment, ART performance 
is quantified  by evaluating   the expected  Kolmogorov-Smirnov statistic for the cumulative distribution function  $F_{\varphi^\ast}(x)$ 
\begin{equation}\label{eq:expKS}
\frac{1}{R}\sum_{r=1}^R \sup_{x}|F_{\varphi,r}^{N,H}(x)-F_{\varphi^\ast}(x) |,
\end{equation}
as a function of the expected computational cost  \eqref{eq:expCost} detailed below.

\subsubsection{Expected Cost}
In both problem \#RE and \#BI (Section \ref{sec:Prob2} and \ref{sec:Prob1}) the expected cost  (in units of the cost of evaluating $S^\ast$) is approximated by
\begin{equation}\label{eq:expCost}
	K+ \frac{1}{R}\sum_{r=1,k=1}^{R,K}N^{(k)}_r \textrm{gain}^{(k)}
\end{equation}
with $ \textrm{gain}^{(k)}$ representing the ratio of the cost of evaluating the reduced score $\Srom^{(k)}$ and {error $\Err^{(k)}$} to the cost of the true score $\Score$, and where $N^{(k)}_r$ represents the number of evaluations of the reduced score at the $k$-th iteration and $r$-th execution of the ART algorithm. 
{Note that this expected cost neglects the extra-cost of updating the reduced model as compared to evaluating the true model. The gain is computed in practice empirically in an off-line phase.} 
{We provide in Appendix~\ref{app:ArthighDim} more results for Model \#RB using CPU time as a reference. For $d=4$, CPU time exhibits very mild discrepancy as compared to~\ref{eq:expCost} and RB update is indeed computationally negligible. This is however no longer true as $d$ increases (around $d=25$).
}

\subsection{Numerical results}

\subsubsection{Experimental setup for evaluation}\label{sec:expSetup}
ART  is configured as follows for the different problems and models.  
The mutation proposal is based on an Ornstein-Uhlenbeck~\cite{garthwaite2016adaptive} adaptive state movement. 

For each experiment, we give in the table below: {the entropic log cost threshold $c_1$ triggering snapshot sampling, the adaptive tempering parameter $c_2$, the total number of snapshots $K$, the parameter $j_0$ triggering non-penalized IS of snapshots,  the number $M$ of Markov transitions at each mutation step,  the number $R$ of runs  and the number  $N$ of particles.} 

\begin{tabular}{l|ccccccccc}
\hspace{-1cm} & {\footnotesize $h$} & {\footnotesize $d$}& {\footnotesize $K$}&{\footnotesize $j_0$}&{\footnotesize$M$} & {\footnotesize $R$} &{\footnotesize $N \in$}& {\footnotesize $c_1$}& {\footnotesize $c_2$}\\
\hline
\hspace{-1cm} {\footnotesize problem~\#RE \& model~\#RB}& 5101& 4 & 200 & 5 & 30 & 50 & [1e2, 3e3] &1e-3 &1e-2 \\
\hspace{-1cm} {\footnotesize problem~\#RE \& model~\#RB}& 5101& 25& 200 & 50 & 30 & 20 & \{1e2\} &1e-1 &1e-2 \\
\hspace{-1cm} {\footnotesize problem~\#RE  \& model~\#S} & 1 & 1 & 200 & 5 & 100 & 100 &  [1e3, 1e4]& 1e-3 &1e-3 \\
\hspace{-1cm} {\footnotesize problem~\#BI  \& model~\#RB} & 5101& 4  & 200 & 5 & 30 & 100 &  [1e1, 5e2]& 1e-3 &1e-2  
\end{tabular}\medskip

For problem~\#BI (which uses model~\#RB), we draw uniformly on the square $\Omega$ a set $\mathcal{Z}$ of $20$ random points $z$,  where we observe the solution $\Psi_z^\ast$ up to a Gaussian noise of variance $1e-2$. The observations are displayed in Figure~\ref{fig:d}.

For problems \#RE (resp. for problem~\#BI), we evaluate the empirical relative square error  \eqref{eq:expError} (resp. the expected  Kolmogorov-Smirnov statistic \eqref{eq:expKS}) versus the average cost and compare to results obtained for several dozen of runs  with the true score $S^\ast$ and for a varying number of particles, using either  the standard adaptive multilevel splitting (AMS) algorithm~\cite{cdfg} (except for problem~\#BI, which cannot be solved by AMS) or  adaptive SMC simulation~\cite{Beskos2016}. In the case of problem~\#RE, we also compare to the performance for  several dozen of  runs of the adaptive reduced multilevel splitting (ARMS) algorithm~\cite{cerou2024adaptive} with a varying number of particles. {Under our experimental protocol, the number of evaluation of the true model in the reference multilevel splitting algorithms (AMS, ARMS) is typically of order $N \times 10^{3}$. For $N=100$, this yields around $500$ times more than in the ART case.}

\subsubsection{Rare event simulation (Problem~\#RE)}

\begin{figure}[ht]
\begin{center}
\begin{tabular}{cc}
\hspace{-2cm}\includegraphics[width=0.6\textwidth]{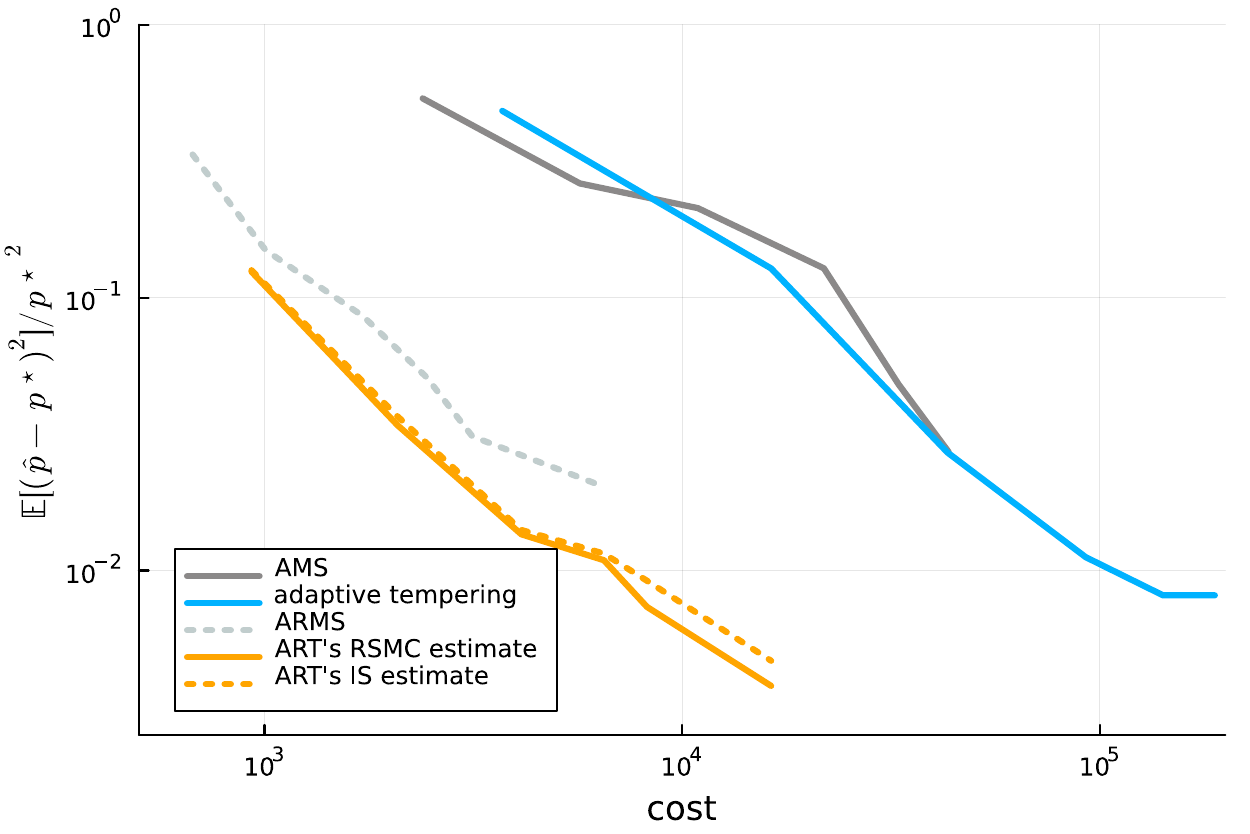}&\includegraphics[width=0.6\textwidth]{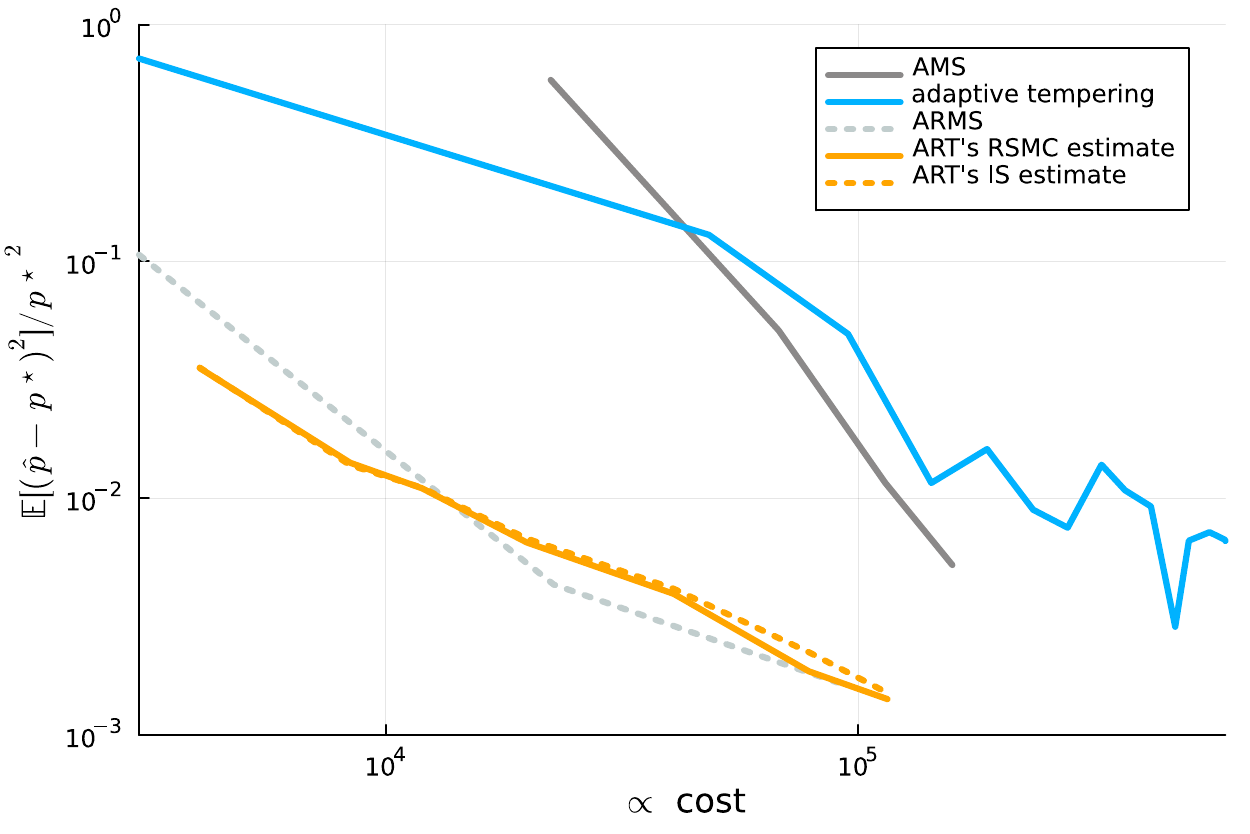}
\end{tabular}
	\caption{{\footnotesize  {  \textbf{Problem~\#RE (rare event simulation) \& model~\#RB (PDE-based score, $d=4$)}.  Relative  expected  square error as a function of the expected cost for ART's IS  and RSMC  estimates,  for a score function \eqref{eq:scoreTheta} with $p=1$ (left) and $p=\infty$ (right), and  for a varying number of particles $N$. Comparison to the performance of  adaptive tempering  with $\Score$ and to  results of AMS  and  ARMS computed in~\cite{cerou2024adaptive}.}}\label{fig:a}}
		\end{center}\vspace{-0.cm}
\end{figure}
\begin{figure}[ht]
\begin{center}
\includegraphics[width=0.6\textwidth]{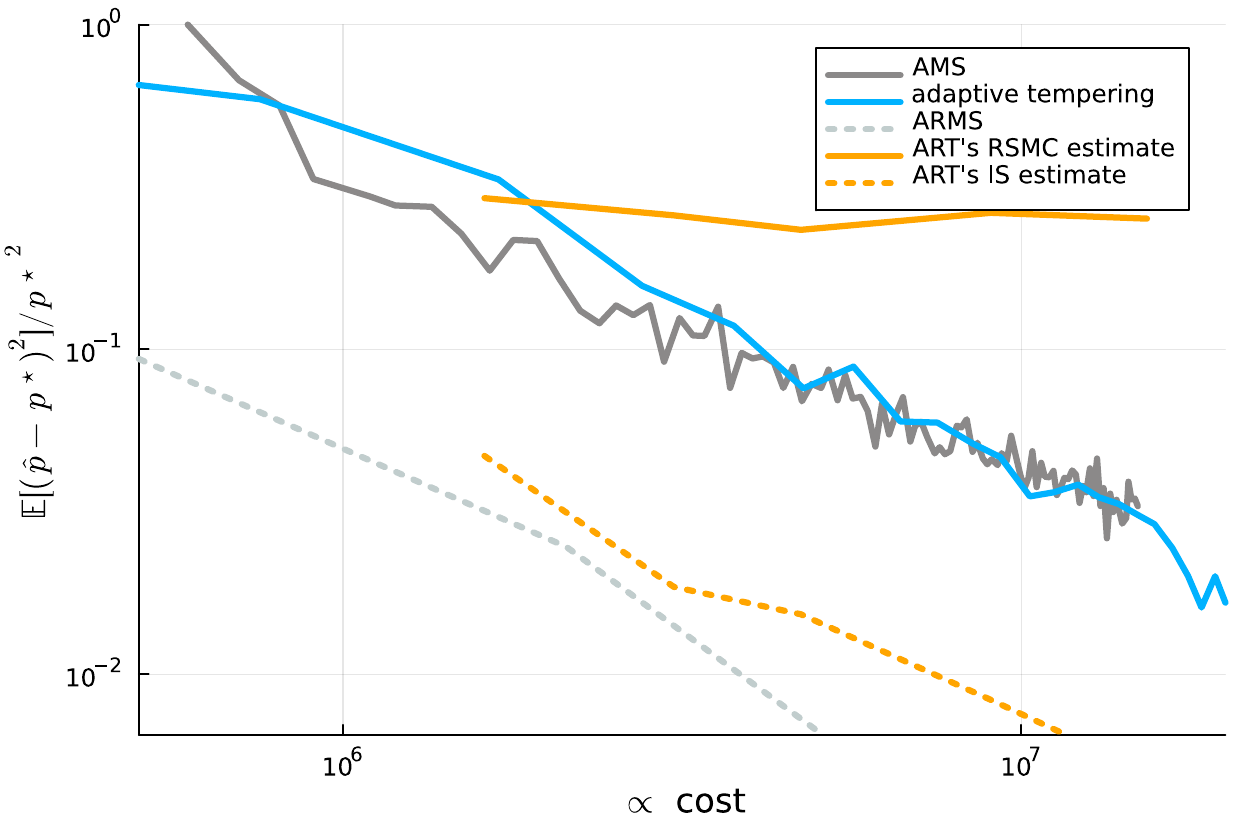}
	\caption{{\footnotesize  {  \textbf{Problem~\#RE (rare event simulation) \& model \#S (closed-form score)}. Relative  expected  square error as a function of the expected cost for ART's IS  and RSMC  estimates for a varying number of particles $N$. Comparison to the performance of  adaptive tempering   with $\Score$  and to  results of AMS  and  ARMS computed in~\cite{cerou2024adaptive}.} \label{fig:b}}}
		\end{center}\vspace{-0.cm}
\end{figure}

Figure~\ref{fig:a}  and ~\ref{fig:b} compare the performances of the ART algorithm with those of the state-of-the-art methods in the case of small dimension of the parameter space:  model~\#RB with $d=4$  or the one-dimensional model~\#S. 

We note that the  reduced sampling methods (ARMS and ART's IS estimate) share similar performances in the case of our rare event simulations. We nevertheless remark that depending on the model, there exists a slight gain or loss of the order of a few units for ART in comparison to ARMS (a gain in the case of model~\#RB for $p=1$ and a loss in the case of model \#S). Discriminating the most performant reduced method (ART or ARMS) is not obvious as performances appear in our experiments to be  model dependent and sensitive to the setting of the free parameters (given in table above). However, we emphasize that the ART algorithm has the advantage of being simpler to implement as it does not require the use of the nesting and domination conditions of the ARMS algorithm. Furthermore, unlike ARMS, to maintain consistency (at least in an idealized case) ART only requires the use of an approximate error bound in cases where a rigorous one is not available (\eg in the case of model~\#RB with $p=\infty$).  
 Last but not least,  the performances of the two reduced sampling methods are far beyond the standard ones  in terms of square error  times cost: we observe  a gain of the order of a decade or two of the expected square error for a given cost.

Although ART's RSMC and IS estimates are almost indistinguishable for model~\#RB, we note that ART's RSMC estimate (which relies solely on the reduced score) deteriorates in the case of model~\#S, the square error tending to be constant with respect to the cost. For this case, using the spline approximation $\Srom$ rather than the true score $\Score$ induces a significant bias. Indeed, the non-differentiability of $\Score$ at the rare event boundary leads to high-frequency oscillations (which tend to decrease with the number of interpolation points, but never disappear completely) in the spline model. The consequence is that a significant proportion of particles in the rare event set $\{x\in \mathcal{X} :\Score(x) \ge 1\}$ is excluded from the set $\{x\in \mathcal{X} :\Srom(x) \ge 1\}$, leading to a negative bias.    On the contrary, we  observe that ART's IS estimate is not biased and retains its good performance in this case, independently of the problem of convergence of $\Srom$ to $\Score$.

On the other hand for both reference estimates (AMS and tempering) and IS estimates for ARMS and ART, we observe approximately a unitary negative slope (on a logarithmic scale) of the variance as a function of cost, implying that the product of variance and cost is constant.   Such a behavior suggests asymptotic normality of the proposed IS estimate, as it has been demonstrated for AMS~\cite{cerou:hal-01417241}.

\subsubsection{Bayesian inversion (Problem~\#BI)}

\begin{figure}[b!]
\begin{center}
\begin{tabular}{c}
\includegraphics[width=0.7\textwidth]{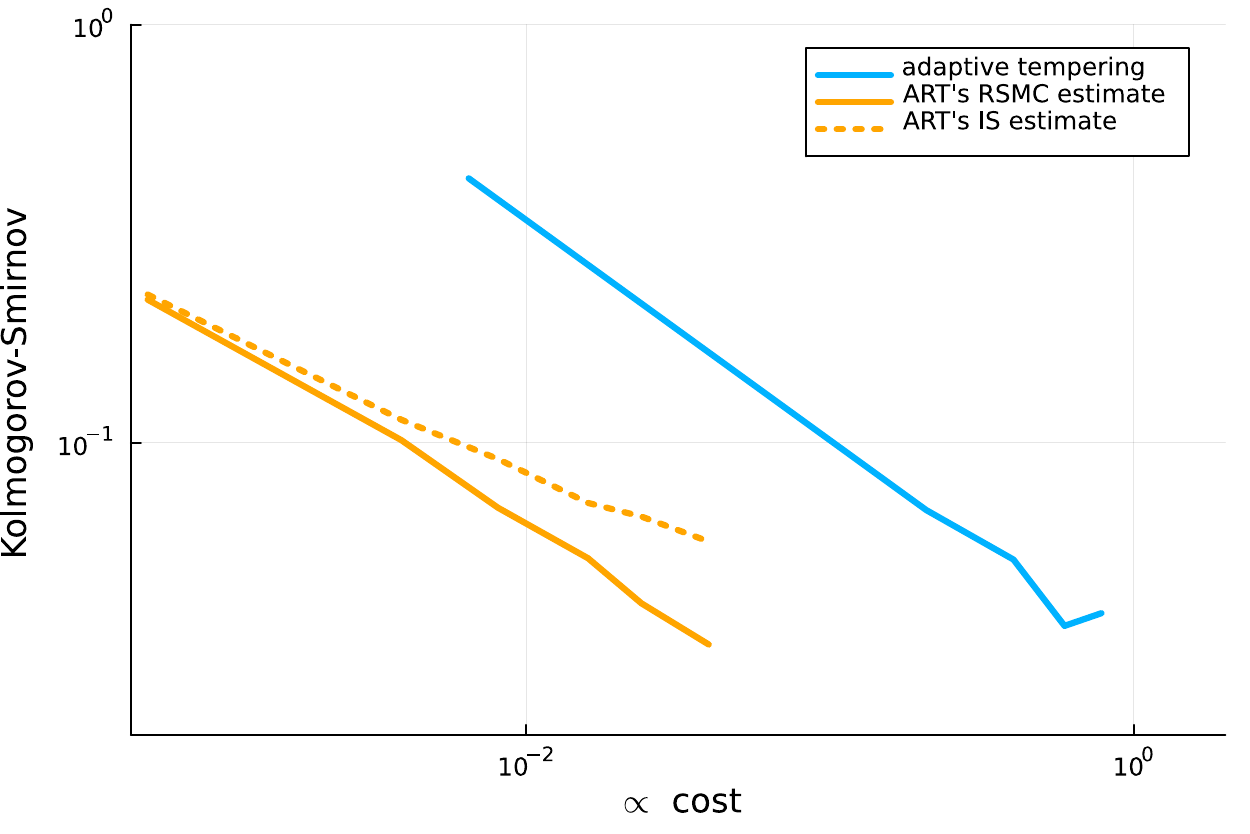}\vspace{-0.35cm}
\end{tabular}
	\caption{{\footnotesize  { {\bf Problem~\#BI (Bayesian inversion) \& model~\#RB (PDE-based score)}.    Kolmogorov-Smirnov distance as a function of the expected cost for ART's IS  and RSMC  estimates  for a varying number of particles $N$. Comparison to the performance of  adaptive tempering  with $\Score$. \label{fig:e}}}}
		\end{center}\vspace{-0.cm}
\end{figure}

We next turn to the Bayesian inverse problem. Figure~\ref{fig:e} shows that the ART algorithm decreases the cost of simulation in comparison to adaptive tempering, of about a decade to obtain a similar sampling accuracy  according to the Kolmogorov-Smirnov distance \eqref{eq:expKS}.   

We observe a sub-unitary negative slope (of minus one half on a logarithmic scale)  of this distance as a function of cost for ART's RSMC estimate, implying that the product of the  square of the Kolmogorov-Smirnov distance and the cost is constant, and suggesting that the product of the estimator's variance and the cost is constant for any test function.  Again this behavior supports the conjecture of asymptotic normality for the proposed estimate. 

In contrast to the case of problem \#RB with model \#S, we find that ART's IS estimate performs slightly worse than ART's RSMC estimate. Indeed, its slope is less than the slope  minus one half  observed for the RSMC estimate.  In this case, we conjecture that for sufficiently large $N$ and a sufficiently accurate approximation $\Srom$, the bias of the ART's IS and RSMC normalized estimates are negligible in comparison to their variances, and furthermore, the variance of ART's IS estimate will be greater than that of ART's RSMC estimate. 
 Indeed, let's assume that $\Srom^{(k)}\simeq \Score$ for iterations $k>K_{j_0}$ (which is a plausible assumption based on the illustrations in Figures~\ref{fig:3} and \ref{fig:3a}).  The RSMC estimate  should then have a variance of at best $1/{NH}$~\cite[Prop. 9.5.6]{del2013mean} and a negligible  bias~\cite{del2007sharp}. On the other hand, based on the Appendix~\ref{app:martingal}, the IS estimate   should have a variance  of the order of $1/{H}$ and a negligible  bias.  Consequently,  considering $N$  tending to infinity and $H$ fixed, the mean square error of the RSMC should tend towards zero with a speed at best inversely proportional to the cost \eqref{eq:expCost} (of the order of $KN$), while the mean square error of the IS estimate will tend at best towards a constant (of the order of $1/H$).\medskip
 
 \subsubsection{Model \#RB in higher dimension}
 
 The core of the ART algorithm relies on sequential Monte Carlo methods for the tempering and bridging routines. Such methods are known to be stable in high dimensions~\cite{Beskos2014}. However, in the specific case of Model \#RB, the computational cost of ART increases substantially with the parameter space dimension $d$. More precisely, updating RB error approximations or {evaluating this RB error} induces a quadratic computational cost  in  $d$,  inherent to the RB model and not the ART algorithm itself. It should be noted that the cost increase {for update} may be acceptable  in the RB formalism as the reduced basis assembly is intended to be  performed off-line. Consequently, as shown in Appendix~\ref{app:ArthighDim}, increasing $d$ from $4$ to $25$   impacts ART  performance with a speed loss of about two decades for a given expected square error, making the method non-competitive. We should nevertheless mention that there are possible remedies to avoid this slowdown with respect to $d$. Indeed, to avoid  excessive cost of RB updates not affordable on-line, recent works challenge the RB off-line/on-line paradigm. In particular, the adaptive reduced modeling scheme of \cite{peherstorfer2015online} proposes inexpensive on-line updates to the RB basis while maintaining its dimension. Other popular works on {\it active subspace} offer ways to circumvent the problem by learning a low-dimensional representation of the parameter space \cite{constantine2014active}.
 
 \subsubsection{Summary}
 In summary, our numerical experiments have demonstrated that a computational cost reduction of over a decade is achieved with the ART algorithm, compared with state-of-the-art standard approaches. In addition, the product of variance and cost appears to be constant for these estimates in some regimes, suggesting asymptotic normality. 
 
 The results also show that, for a particular problem (Problem~\#S), the RMSC estimate can be biased even in the case where the reduced score tends towards the true score. Conversely, on the Bayesian inverse problem, the RSMC estimate seems to perform slightly better than the IS estimate for a fixed budget of snapshots. 

\section{Conclusions and Perspectives}

This work proposes an adaptive algorithm called ART that reduces the cost of adaptive SMC simulation when the  score function $\Score$ is very expensive to compute. 

The general idea of ART is to  perform iterative importance sampling of a sequence of target distributions of the form $\propto  e^{\beta^{(k)}{\Score}} d\pi$ parametrized   at iteration $k$ by the inverse temperature  $\beta^{(k)} \in \mathbb{R}_+$, using a sequence of proposals  of the form  $\mu_{\beta^{(k)}}^{(k)} \propto e^{\beta^{(k)}{\Srom^{(k)}}} d\pi$, much cheaper to simulate with SMC. As the algorithm is iterated, the proposals are adapted by adjusting the inverse temperature $\beta^{(k)}$ and refining the approximation of the reduced score $\Srom^{(k)}$ using the history of score evaluations previously calculated for importance sampling. 

As $\Srom^{(k)}$ gains in accuracy and $\beta^{(k)}$ increases, the reduced score approximation should be refined using samples of the state space in regions closer and closer to the low-temperature event of interest. However, an essential ingredient in achieving this desirable convergence is to determine the first non-achievable inverse temperature $\beta^{(k)}$ suitable for the accuracy of the current reduced score approximation $\Srom^{(k)}$.  Justified by theoretical arguments, this problem is addressed in ART using an entropy constraint: $\beta^{(k)}$ is chosen such that  the relative entropy between a surrogate target and the proposal corresponds to a given logarithmic cost of importance sampling. The surrogate target is constructed using an approximation of the quantification of the worst-case error associated with the reduced score approximation.

A final point addressed by ART is the adaptation of the proposal to a new approximation of the reduced score $\Srom^{(k+1)}$. To achieve this, ART   searches among past simulation indexes $k_{bridge}<k+1$ for an approximation $\Srom^{(k_{bridge})}$ and an initial parameter $\beta_{bridge}(k+1)$ such that  the relative entropy between the new  proposal  $\mu_{\beta_{\mrm{bridge}}(k+1)}^{(k+1)}$ at its initial inverse temperature and the past proposal ${  \mu_{\beta^{(k_{\mrm{bridge}})}}^{(k_{\mrm{bridge}})} }$ at its critical inverse temperature  is kept below a certain standard  threshold and, in addition, such that relative entropy between the  surrogate target and the proposal is below the given logarithmic cost of importance sampling. The algorithm then bridges the two proposals  in a single SMC simulation step. 

In addition to demonstrating their consistency  in an idealized setting, we evaluate in our numerical experiments the empirical convergence of the two proposed estimators, an importance sampling estimator and a reduced SMC estimator, in the context of rare event simulation and  for a Bayesian inverse problem based on a PDE.  We exhibit empirically that we significantly reduce simulation cost of order of a decade while achieving a comparable accuracy.

\appendix

\section{Consistency}\label{app:martingal}
We show the consistency of  the unnormalized IS estimator $\hat{\gamma}_{\beta_\infty}^{H}$  in the idealized setting, see Section~\ref{sec:IS}.\medskip

\begin{Pro}
For any test function $\varphi$, the  estimator $ \hat{\gamma}_{\beta_\infty}^{H} $ is unbiased, \ie 
$
\E \, \hat{\gamma}_{\beta_\infty}^{H}(\varphi)  = \gamma_{\beta_\infty}^\ast (\varphi),
$
and its  relative variance is 
\begin{equation*}
\frac{ \Var\b{  \hat{\gamma}_{\beta_\infty}^{H} (\varphi) }}{{\gamma_{\beta_\infty}^\ast (\varphi)}^2} =\frac{1}{H^2}  \sum_{k=K_{j_0}+1}^{K_{j_0}+H} \frac{\mathbb{E}(Z^{(k)}_{\beta^{(k)}}\pi(e^{2{\beta_\infty \Score} -{{\beta}^{(k)}} \Srom^{(k)} } \varphi^2) )}{{\gamma_{\beta_\infty}^\ast (\varphi)}^2} - 1, 
\end{equation*}
 where the expectation is with respect to the random variable ${{\beta}^{(k)}} \Srom^{(k)}$. 
\end{Pro}\medskip

\proof{
We first show that 
the random sequence
$$
H \mapsto M^\varphi_H \eqdef H \, ( \hat \gamma^H_{\beta_\infty}(\varphi)- \gamma^\ast_{\beta_\infty}(\varphi)),
$$
is a martingale with respect to the filtration generated by the snapshots $\mathcal{F}_H=\sigma(X_1,\ldots,X_{K_{j_0}+1+H})$. Indeed, defining the  density ratio
$$f^{(k)}_{\beta_\infty}(x)=Z^{(k)}_{\beta^{(k)}}e^{\beta_\infty\Score(x) - \beta^{(k)}\Srom^{(k)}(x)},$$
the IS estimator \eqref{eq:unestimator} is rewritten as 
$$  \hat \gamma^H_{\beta_\infty}(\varphi)=\frac{1}{H}\sum_{k=K_{j_0}+1}^{K_{j_0}+H} f^{(k)}_{\beta_\infty}(X_{k+1})  \varphi({X_{k+1}}),
$$
 and we have
\begin{align*}
\mathbb{E}[  M^\varphi_H |  \mathcal{F}_{H-1}]&=\mathbb{E}[  \sum_{k=K_{j_0}+1}^{K_{j_0}+H} f^{(k)}_{\beta_\infty}(X_{k+1}) \varphi({X_{k+1}}) - H\gamma^\ast_{\beta_\infty}(\varphi) | \mathcal{F}_{H-1}],\\
&=M^\varphi_{H-1}+\mathbb{E}[ f^{({K_{j_0}+H})}_{\beta_\infty}(X_{{K_{j_0}+H}+1}) \varphi({X_{{K_{j_0}+H}+1}}) |  \mathcal{F}_{H-1}] - \gamma^\ast_{\beta_\infty}(\varphi),\\
&=M^\varphi_{H-1},										 
\end{align*}
where the last equality is due to the fact that  the snapshot $X_{{K_{j_0}+H}+1}$ is drawn according to the proposal  $$\mu^{({K_{j_0}+H})}_{\beta^{({K_{j_0}+H})}}=e^{\beta^{({K_{j_0}+H})}\Srom^{({K_{j_0}+H})}}d\pi /Z^{({K_{j_0}+H})}_{\beta^{({K_{j_0}+H})}}.$$
Now, since $
M^\varphi_H $
is a martingale, the estimator is unbiased $\E \, \hat{\gamma}_{\beta_\infty}^{H} (\varphi) = \gamma_{\beta_\infty}^\ast (\varphi),$ and its variance is given by the average of the quadratic variations of the sequence. The ${K_{j_0}+H}$-th element of the sequence, the expectation of the quadratic variation  is  (in this ideal setup the normalization constant can be computed exactly given some $\beta^{(k)}$ and $\Srom^{(k)}$)
 $$
\Var_{\mu^{({K_{j_0}+H})}_{{\beta}^{({K_{j_0}+H})}}}(\frac {d\gamma_{\beta_\infty}^\ast}{d\mu^{({K_{j_0}+H})}_{{\beta}^{({K_{j_0}+H})}}} (X)\varphi(X) | \mathcal{F}_{H-1})  = Z^{(k)}_{\beta^{(k)}} \pi(e^{2{\beta_\infty \Score} -{{\beta}^{(k)}} \Srom^{(k)} } \varphi^2) - {\gamma_{\beta_\infty}^\ast(\varphi)}^2,
 $$ 
where ${{\beta}^{(k)}} \Srom^{(k)} $ is a variable depending on the randomness of the  previous snapshots.  Therefore,  the  variance is
\begin{align*}
{ \Var\b{  \hat{\gamma}_{\beta_\infty}^{H} (\varphi) }}
&=\frac{1}{H^2}  \sum_{k=K_{j_0}+1}^{K_{j_0}+H}  \mathbb{E}( Z^{(k)}_{\beta^{(k)}}\pi(e^{2{\beta_\infty \Score} -{{\beta}^{(k)}} \Srom^{(k)} } \varphi^2)) - {{{\gamma_{\beta_\infty}^\ast (\varphi)}^2}}. \quad \square
\end{align*}

}

\section{ART \& RB in high dimensions}\label{app:ArthighDim}
We show in this appendix that increasing the parameter space dimension $d$ of the RB model (while keeping the other parameters in an analogous setting, see the table in Section~\ref{sec:expSetup}), yields non-competitive results. The main reason for this performance drop is a quadratic complexity with respect to $d$: an RB update (with error) scales in $\mathcal{O}(hd^2K^2)$), and an error evaluation  scales as $\mathcal{O}(d^2K^2)$. As a consequence, the following table\medskip
	\begin{center}
\begin{tabular}{c|c|c|c|c|c|c|c|c|}
	\hspace{-1cm}  &  \multicolumn{2}{c|}{CPU } & \multicolumn{1}{c|}{CPU$\times$ \eqref{eq:expError}} & \multicolumn{1}{c|}{\eqref{eq:expCost}$\times$\eqref{eq:expError}} \\
	\hline 
	\hspace{-1cm} &   \multicolumn{1}{c|}{$\Srom(x)$ {\it v.s.} $\Score(x)$ } &\multicolumn{1}{c|}{ $\Srom(x)$ \& \Err(x) {\it v.s.} $\Score(x)$ }& \multicolumn{1}{c|}{ART {\it v.s.}  SMC } &  \multicolumn{1}{c|}{ART {\it v.s.} SMC }  \\
	\hline 
	\hspace{-1cm}$d$=4& $\times 63$& $\times 29$ &$\times 17$&$\times 27$\\
	\hline 
	\hspace{-1cm}$d$=25& $\times 12$ & $\times 3$ &$\times 1$&$\times 2$
\end{tabular}
\end{center}\medskip
shows that increasing $d$ slows the ART algorithm and thus deteriorates its performance:  while for $d=4$ the speed of the ART algorithm is for a given expected square error \eqref{eq:expError} nearly 20 times faster than adaptive SMC in terms of CPU time and more than 20 times less expensive in terms of the cost  \eqref{eq:expCost}, both algorithms share comparable performance for $d=25$.
A closer analysis of the previous table shows that increasing $d$  slows the relative cost (as compared to a true model evaluation) of solving an RB system (scaling in $\mathcal{O}(K^3+dK^2)$) by a factor of 6, and adding the cost of  error estimation (scaling in $\mathcal{O}(d^2K^2)$) lowers  this speed by up to a factor  10.
More importantly,  for $d=25$, the cost of  RB updates becomes critically high even if rare, and exceeds the cost of the many evaluations of the RB solution and  its error or the cost of computing snapshots, as detailed in the next table:\medskip

\begin{tabular}{c|c|c|c|c|c|}
\hspace{-1cm}&  RB update  &   {tempering} & {bridging} & {$S^\ast$ eval.} \\
\hline 
\hspace{-1cm} \%CPU time, $d$=4&  { 3.3\%} & 63.7\% (2.5\% optim.)&32.0\%  (14.9\% optim.) & 1.0\% \\
 \hline 
\hspace{-1cm} \%CPU time, $d$=25& { 68.4\%} & 15.4\% ($<$0.1\% optim.) &16.1\%  (0.4\% optim.)& 0.1\% 
\end{tabular}\medskip 

We stress that the performance drop is due the computational complexity of the RB update and evaluation, especially of the error and does not come from the ART algorithm methodology itself.

\bibliographystyle{plain}
\bibliography{mathias}

\end{document}